\documentclass[twoside,a4paper]{IEEEtran}

\usepackage[noadjust]{cite} 
\usepackage{multirow}
\usepackage{amsmath}
\usepackage{overpic}
\usepackage{amsopn}
\usepackage{amssymb}
\usepackage{lettrine}
\usepackage{accents}
\usepackage{mathrsfs}
\usepackage{pifont}
\usepackage{setspace}
\usepackage{stackengine}
\newcommand\xrowht[2][0]{\addstackgap[.5\dimexpr#2\relax]{\vphantom{#1}}}
\usepackage{bm}         
\usepackage{fix-cm}  
\usepackage{algorithm}
\usepackage{algorithmicx}
\usepackage{algpseudocode}

\makeatletter
\newcommand{\removelatexerror}{\let\@latex@error\@gobble}
\makeatother

\usepackage{mleftright} 
\mleftright                 

\usepackage{mathdots}   

\makeatletter
\def\IEEElabelanchoreqn#1{\bgroup
\def\@currentlabel{\p@equation\theequation}\relax
\def\@currentHref{\@IEEEtheHrefequation}\label{#1}\relax
\Hy@raisedlink{\hyper@anchorstart{\@currentHref}}\relax
\Hy@raisedlink{\hyper@anchorend}\egroup}
\makeatother

\usepackage[utf8]{inputenc} 
\usepackage[T1]{fontenc} 

\usepackage{tikz}
\usepgflibrary{arrows.meta}
\usetikzlibrary{positioning}
\usetikzlibrary{shadows} 
\usetikzlibrary{calc}
\usetikzlibrary{patterns}
\usepackage{pgfplots}
\pgfplotsset{compat=1.15}
\usepgfplotslibrary{units}
\usetikzlibrary{spy,backgrounds}
\usetikzlibrary{decorations.markings}
\usepackage{pgfplotstable}
\usetikzlibrary{positioning, fit, calc} 
\tikzset{block/.style={draw, thick, minimum width=0.5cm, minimum height=0.5cm, align=center}, 
	line/.style={-latex}   
}
\newtheorem{theorem}{Theorem}
\newtheorem{lemma}{Lemma}
\newtheorem{corollary}{Corollary}

\newtheorem{proposition}{Proposition}

\newtheorem{remark}{Remark}
\newtheorem{definition}{Definition}
\newtheorem{example}{Example}

\newcommand{\remarksymbol}{\text{$\blacktriangle$}}

\usepackage{hyphenat} 

\newcommand{\expec}[1]{\mathbb{E}{\left[#1\right]}}

\newcommand{\midk}[1]{\kern0.1em #1 \kern0.1em}
\newcommand{\middlek}[1]{\kern0.1em \middle#1 \kern0.1em}
\newcommand{\bigk}[1]{\kern-0.1em \bigm#1 \kern-0.1em}
\newcommand{\Bigk}[1]{\kern-0.1em \Bigm#1 \kern-0.1em}
\newcommand{\biggk}[1]{\kern-0.1em \biggm#1 \kern-0.1em}
\newcommand{\Biggk}[1]{\kern-0.1em \Biggm#1 \kern-0.1em}

\newcommand{\tn}[1]{\textnormal{#1}}

\newcommand{\vect}[1]{\mathbf{#1}} 

\newcommand{\II}{\mathop{}\!\const{I}}
\newcommand{\mat}[1]{\mathbb{#1}}
\newcommand{\trans}[1]{#1^{\textnormal{\textsf{\tiny T}}}} 
\newcommand{\trace}[1]{\operatorname{tr}\left(#1\right)} 

\newcommand{\diag}[1]{\mathsf{diag}\left(#1\right)} 

\newcommand{\const}[1]{\textnormal{\usefont{U}{eur}{m}{n}\selectfont #1}} 
\newcommand{\hh}{\mathop{}\!\const{h}}  
\newcommand{\relD}{\mathop{}\!\mathsf{D}}         
\newcommand{\relDf}[2]{\relD\left(#1 \kern0.1em\middle\|\kern0.1em #2\right)}
\newcommand{\erelDf}[2]{\relD(#1 \kern0.1em\|\kern0.1em #2)} 
\newcommand{\bigrelDf}[2]{\relD\bigl(#1 \kern-0.1em \bigm\| \kern-0.1em#2\bigr)}
\newcommand{\BigrelDf}[2]{\relD\Bigl(#1 \kern-0.1em \Bigm\| \kern-0.1em#2\Bigr)}
\newcommand{\biggrelDf}[2]{\relD\biggl(#1 \kern-0.1em \biggm\| \kern-0.1em#2\biggr)}
\newcommand{\BiggrelDf}[2]{\relD\Biggl(#1 \kern-0.1em \Biggm\| \kern-0.1em#2\Biggr)}
\newcommand{\overbar}[1]{\mkern 1.5mu\overline{\mkern-1.5mu#1\mkern-1.5mu}\mkern 1.5mu}
\newcommand{\nt}{n_{\tn{T}}}
\newcommand{\nr}{n_{\tn{R}}}

\newcommand{\Prvcond}[2]{\Pr\left[#1 \kern0.1em\middle|\kern0.1em #2\right]}
\newcommand{\ePrvcond}[2]{\Pr[#1 \kern0.1em|\kern0.1em #2]} 
\newcommand{\bigPrvcond}[2]{\Pr\bigl[#1 \kern-0.1em \bigm| \kern-0.1em#2\bigr]}
\newcommand{\BigPrvcond}[2]{\Pr\Bigl[#1 \kern-0.1em \Bigm| \kern-0.1em#2\Bigr]}
\newcommand{\biggPrvcond}[2]{\Pr\biggl[#1 \kern-0.1em \biggm| \kern-0.1em#2\biggr]}
\newcommand{\BiggPrvcond}[2]{\Pr\Biggl[#1 \kern-0.1em \Biggm| \kern-0.1em#2\Biggr]}

\newcommand{\Prscond}[2]{\Pr\left(#1 \kern0.1em\middle|\kern0.1em #2\right)}
\newcommand{\ePrscond}[2]{\Pr(#1 \kern0.1em|\kern0.1em #2)} 
\newcommand{\bigPrscond}[2]{\Pr\bigl(#1 \kern-0.1em \bigm| \kern-0.1em#2\bigr)}
\newcommand{\BigPrscond}[2]{\Pr\Bigl(#1 \kern-0.1em \Bigm| \kern-0.1em#2\Bigr)}
\newcommand{\biggPrscond}[2]{\Pr\biggl(#1 \kern-0.1em \biggm| \kern-0.1em#2\biggr)}
\newcommand{\BiggPrscond}[2]{\Pr\Biggl(#1 \kern-0.1em \Biggm| \kern-0.1em#2\Biggr)}

\newcommand{\Exp}{\operatorname{\textnormal{\mathsf{E}}}}

\newcommand{\Econd}[3][]{\Exp_{#1}\left[#2 \kern0.1em\middle|\kern0.1em #3\right]}
\newcommand{\eEcond}[3][]{\Exp_{#1}[#2 \kern0.1em|\kern0.1em #3]}
\newcommand{\bigEcond}[3][]{\Exp_{#1}\bigl[#2 \kern-0.1em \bigm| \kern-0.1em #3\bigr]}
\newcommand{\BigEcond}[3][]{\Exp_{#1}\Bigl[#2 \kern-0.1em \Bigm| \kern-0.1em #3\Bigr]}
\newcommand{\biggEcond}[3][]{\Exp_{#1}\biggl[#2 \kern-0.1em \biggm| \kern-0.1em #3\biggr]}
\newcommand{\BiggEcond}[3][]{\Exp_{#1}\Biggl[#2 \kern-0.1em \Biggm| \kern-0.1em #3\Biggr]}

\newcommand{\dd}{\mathop{}\!\mathrm{d}}

\newcommand{\supp}{\operatorname{supp}}

\makeatletter

\newcommand{\Rmnum}[1]{\expandafter\@slowromancap\romannumeral #1@}
\makeatother

\usepackage{threeparttable}
\usepackage{diagbox}
\usepackage{array}
\usepackage{boldline}
\usepackage{caption}
\usepackage{dsfont} 

\usepackage{algpseudocode}

\allowdisplaybreaks
\usetikzlibrary{datavisualization} 

\usetikzlibrary{spy,backgrounds}
\usetikzlibrary{decorations.markings}

\begin{document}

\title{Capacity Results for Multiple-Input Multiple-Output Optical Wireless Communication With Per-Antenna Intensity Constraints}

\author{Ru-Han Chen, Longguang Li$^\dagger$, Jia-Ning Guo, Xu Yang, Jian Zhang, and Lin Li
	\thanks{This work is supported in part by the National Natural Science Foundation of China under Grant No. 62071489 and 62101192, and in part by Shanghai Sailing Program under Grant No. 21YF1411000. \textit{(Corresponding author: Longguang Li.)}}
	\thanks{R.-H. Chen and X. Yang are with Sixty-third Research Institute, National University of Defense Technology, Nanjing, China (e-mail: tx\_rhc22@nudt.edu.cn, fractal\_yangxu@outlook.com).}
	\thanks{J. Zhang, J.-N. Guo and L. Li are with National Digital Switching System Engineering and Technological Research Center, Zhengzhou, China (e-mail: zhang\_xinda@126.com, 14291003@bjtu.edu.cn, wclilin@163.com).} 
	\thanks{L. Li is with Dept. Communication and Electronic Engineering, East China Normal University, Shanghai, China  (e-mail: lgli@cee.ecnu.edu.cn). 
}}

\maketitle

\begin{abstract}
In this paper, we investigate the capacity of a multiple-input multiple-output (MIMO) optical intensity channel (OIC) under per-antenna peak- and average-intensity constraints. We first consider the case where the average intensities of input are required to be equal to preassigned constants due to the requirement of illumination quality and color temperature. When the channel graph of the  MIMO OIC is strongly connected, we prove that the strongest eigen-subchannel must have positive channel gains, which simplifies the capacity analysis. Then we derive various capacity bounds by utilizing linear precoding, generalized entropy power inequality, and QR decomposition, etc. These bounds are numerically verified to approach the capacity in the low or high signal-to-noise ratio regime. Specifically, when the channel rank is one less than the number of transmit antennas, we derive an equivalent capacity expression from the perspective of convex geometry, and new lower bounds are derived  based on this equivalent expression. Finally, the developed results are extended to the more general case where the average intensities of input are required to be no larger than preassigned constants.

%
%
\end{abstract}

\begin{IEEEkeywords}
Channel capacity, dimming control, intensity-modulation and direct-detection (IM/DD), multiple-input multiple-output, optical wireless communication, per-antenna constraint, visible light communication.
\end{IEEEkeywords}

\section{Introduction}
\label{sec:introduction}
\lettrine[lines=2]{W}{ith} 
an ever-growing demand for wireless data connectivity, there is an urgent need for significantly higher capacity and energy efficiency of the existing communication network \cite{you2021towards}. 
As one enabler for development of future communications, optical wireless communication (OWC) utilizes optical spectrum including infrared, visible light, and ultraviolet bands. In recent years, the OWC has attracted significant interest from both academia and industry for its notable advantages, such as vast license-free bandwidth, resistance to radio-frequency (RF) interference, low-cost deployment, inherent secrecy, and integrated lighting, communication and positioning, and been deemed as a complementary technology to the RF counterpart to alleviate the RF spectrum crunch \cite{Chi2015,Elgala2011Indoor}. 

Instead of the coherent transmission adopted in RF communications, the intensity modulation/direct detection (IM/DD) is widely used in most practical OWC systems due to less complexity and lower cost. In an IM/DD mode, information is carried on the modulated instantaneous optical intensity emitted from the transmitters, and the receivers use photodiodes (PDs) to measure the intensity of incoming light. For this reason, the intensity-modulated signal should be real and nonnegative. In addition, the peak and the average intensities of input signals are usually constrained as well from considerations of eye safety, hardware limitations, illumination adjustment, and energy efficiency \cite{chaaban2020capacity}. Due to the above-mentioned constraints on nonnegativity, amplitude, or first-order moment of the input signal, completely characterizing the fundamental limit of the optical intensity channel (OIC) is difficult. Instead, many capacity bounds and asymptotics have been found for the OIC under different input constraints  \cite{chaaban2020capacity,lapidothmoserwigger09_7,faridhranilovic09_1,faridhranilovic10_1,chaaban2016FSO,chaabanrezkialouini17_1,chaabanrezkialouini18_1,limoserwangwigger20_1,moserwangwigger18_3,chaabanrezkialouini18_2,chen2022MISO}. Early works mainly focused on the single-input single-output (SISO) OICs \cite{lapidothmoserwigger09_7,faridhranilovic09_1,faridhranilovic10_1,chaaban2016FSO}. In \cite{lapidothmoserwigger09_7}, entropy power inequality (EPI)-based capacity lower bounds, duality-based capacity upper bounds and asymptotic results are derived for the SISO OIC under a peak-intensity constraint, an average-intensity constraint, and both, respectively. Afterwards, some refined capacity bounds have been derived in~\cite{faridhranilovic09_1,faridhranilovic10_1,chaaban2016FSO} via sphere-packing or slightly different duality-based arguments. Similarly capacity results have been also derived for multiple-input single-output (MISO) and multiple-input multiple-output (MIMO) OICs~\cite{moserwangwigger18_3,chaabanrezkialouini17_1,chaabanrezkialouini18_1,limoserwangwigger20_1}. 


It should be noted that in the above mentioned multi-antenna channels, the existing works usually impose the average-intensity constraint on the total average optical intensity of all transmit antennas. However, such a treatment may not be suitable for the visible light communication (VLC) with fine-grained dimming control and chromaticity adjustment \cite{gancarz2013impact}. In order to achieve high-quality lighting, it is necessary to synthesize multi-color beams to achieve a given color temperature, which requires the optical power of LEDs with different colors to meet a certain ratio \cite{gong2015power}. Another exception is the arisen application of the inter-satellite laser communication networking, where each transmit antenna is geographically separated and has individual Erbium-doped fiber amplifier \cite{khalighiuysal14_1,loyka_17,chan2024optical}. Motivated by the above fact, this paper is concerned with the MIMO OWC system, where arbitrary peak- and average-intensity constraints are individually imposed on each transmit optical antenna. To the best of the authors' knowledge, the relevant results are only found in \cite{chaabanrezkialouini18_2} and our recent paper \cite{chen2022MISO}. In~\cite{chaabanrezkialouini18_2}, the low-SNR capacity slope is characterized for general MIMO OIC with peak intensity constraints and preassigned average intensities. In~\cite{chen2022MISO}, we have investigated the MISO OIC under two types of per-antenna intensity constraints, where both types of channels are proved to be equivalent to SISO OICs under a peak-intensity and several stop-loss moment constraints. The results in \cite{chen2022MISO} relies on stochastic order and a nontrivial decomposition of a nonnegative and bounded random variable, which cannot be directly extended to the linear vector Gaussian channel of our interest.


%

In light of above observations, this paper is concerned with capacity bounds for the MIMO OIC with per-antenna intensity constraints. Our main contributions include
\begin{enumerate}
	\item \textit{Positivity of Strongest Eigen-Subchannel:} 
	We introduce the definition of the channel graph, and prove that the strongest eigen-subchannel of any MIMO OWC channel with a strongly connected channel graph must have strictly positive channel gains; see Thm. \ref{thm:rankone}. This property shows an interesting difference between the OWC and the RF communication in the MIMO setting, and simplifies the capacity analysis.
	
	\item \textit{New Capacity Lower Bounds:} We derive capacity lower bounds of two transceiver frameworks: 1) a certain type of the linear precoding scheme with optimized precoder design (see Props. \ref{prop:linear_precoding} and \ref{prop:linear_precoding2}); 2) the receiver based on the QR decomposition and the successive interference cancellation (see Prop. \ref{prop:ec_oic_qr}).
Another capacity lower bound is derived by using the generalized EPI. We also generalize our results to the case of bounded average intensities, for which a property of the maximum-trace intensity allocation and a near-optimal intensity allocation scheme are presented.
		\item \textit{New Capacity Results on Rank-$(\nt-1)$ Channel:} 
		 For the MIMO OIC of rank $\nt-1$, where $\nt$ optical transmit antennas individually have preassigned average intensities, we equivalently convert it to a vector Gaussian channel with moment constraints from the perspective of convex geometry (see Thm. \ref{thm:nt-1}), based on which the maximum differential entropies of the equivalent input are characterized for the MIMO OICs of rank $\nt-1$ and with two different per-antenna intensity constraints; see Thms. \ref{thm:highsnr_ECMC} and \ref{thm:highsnr_BCMC}. 
\end{enumerate}

The rest of the paper is organized as follows. {\color{black}We end the introduction with notations used throughout the paper.} Sec.~\ref{sec:model} introduces the model of the MIMO OIC with individual average-intensity equality constraints, for which a simplification method is also provided. Sec.~\ref{sec:3} establishes various capacity results for the aforementioned MIMO OIC. Sec.~\ref{sec:equiv} presents an equivalent capacity expression in the special case, by which the effect of the channel matrix can be removed and the original linear model is simplified to a vector Gaussian channel. Sec~\ref{sec:bc-oic} generalizes the capacity results to the case of bounded-cost constraints. Numerical results are exhibited in Sec. \ref{sec:numerical}. The paper is concluded in Sec.~\ref{sec:conclusion}. 

\textit{Notations:} The operator $\|\cdot\|_{p}$ denotes the $\ell_p$-norm of a vector. For any positive integer $n$, the index set $\{ 1,2,\cdots,n\}$ is denoted by $[n]$. For two $n$-dimensional vectors $\bm{\alpha}$ and $\mathbf{b}$, $\mathbf{c}=\max\{\bm{\alpha},\mathbf{b}\}$ denotes the element-wise maximum, i.e., $c_i=\max\{a_i,b_i\}$ for all $i\in [n]$. Given an index set $\mathcal{J}\subseteq [n]$ and a matrix $\mathbb{H}$ with $n$ columns, $\mathbb{H}_{\mathcal{J}}$ denotes the submatrix of entries that lie in the columns indexed by $\mathcal{J}$. For a column vector $\mathbf{x}$, $\diag{\mathbf{x}}$ denotes a diagonal matrix with main diagonal elements $x_1$, $x_2$, $\cdots$, and $x_n$. All logarithms are natural logarithms, and hence all rates specified in this paper are in nats per channel use.



\section{Channel Model}\label{sec:model}
Consider an $\nr \times \nt$ MIMO optical intensity channel with the channel output 
\begin{equation}\label{eqn:model1}
	\mathbf{Y}=\mathbb{H}\mathbf{X}+\mathbf{Z},
\end{equation}
where the nonnegative matrix $\mat{H}\in \mathbb{R}_+^{\nr \times \nt}$ denotes the deterministic channel response between $\nt$ optical transmit antennas (e.g., light-emitting diodes or laser diodes) and $\nr$ optical receive antennas (e.g., PDs); where the $\nr$-dimensional vector $\vect{Z}$ represents the sum of the ambient shot noise induced by the background radiations and the thermal noise, and is assumed to be AWGN, i.e., $
	\vect{Z} \sim \mathcal{N}(\bm{0}_{\nr},\sigma^2\mat{I}_{\nr})$, and independent of $\mathbf{X}$; and  where the vector $\vect{X}=\trans{(X_1, \ldots, X_{\nt})}$ denotes the
	channel input with entries proportional to the optical intensities emitted from $\nt$ transmitters, and hence is nonnegative, i.e.,
	$
	\vect{X} \succcurlyeq \bm{0} $. Due to the requirement of fine-grained adjustment of brightness and chromaticity, or consideration of distributed deployment, the channel input $\vect{X}$ is individually subject to the following per-antenna intensity constraints\footnote{Since the only assumption on channel gains is nonnegativity, we can normalize all maximum allowed peak intensities to unity.} 
	\begin{subequations}\label{eqn:ecc}
		\begin{align}
			&0\le X_k \le 1 ,  \label{eqn:ecc1}\\
			&\mathbb{E}\left[X_k\right] = \alpha_k, ~  \forall k\in [\nt], \label{eqn:ecc2}
		\end{align}
	\end{subequations}
	say, \textit{equal-cost constraints} as discussed in \cite{chen2022MISO}, where the constant $\alpha_k\in \left[0,1\right]$ denotes the ratio of the average intensity to the maximum allowed peak intensity for the $k$-th transmitter. In the remainder of the paper, we refer to the MIMO OIC \eqref{eqn:model1} under the per-antenna constraints \eqref{eqn:ecc} as \textit{the MIMO EC-OIC} for simplicity.

%
%
%
 

\subsection{Channel Reduction}\label{sec:reduction}
Denote the rank of the channel matrix $\mathbb{H}$ by $r\in \mathbb{N}$.
Analogue to \cite{limoserwangwigger20_1}, by using the singular value decomposition  we may equivalently transform the original channel into the following one
\begin{equation}\label{eq:equiv}
	\widetilde{\mathbf{Y}}=\widetilde{\mathbb{H}}\mathbf{X}+\widetilde{\mathbf{Z}},
\end{equation}
where the equivalent channel matrix $
\widetilde{\mathbb{H}}\triangleq\diag{\sigma_1,\cdots,\sigma_r}\cdot\trans{\mathbb{V}_{ 1}} \in \mathbb{R}^{r\times \nt}$ has a full row rank, constants $\sigma_1\ge \sigma_2 \ge  \cdots \ge \sigma_r >0$ are $r$ singular values of $\mathbb{H}$, and the semi-orthogonal matrix $\mathbb{V}_{1}=\left[\mathbf{v}_1,\cdots,\mathbf{v}_r\right]\in \mathbb{R}^{\nt \times r}$ with the column vector $\mathbf{v}_i$ being the eigenvector of $\trans{\mathbb{H}}\mathbb{H}$ with respect to its eigenvalue $\sigma_i^2$, and the AWGN $\widetilde{\mathbf{Z}} \sim \mathcal{N}(\bm{0}_{r},\sigma^2\mat{I}_{r})$.


\begin{remark}\label{remark:2-2}
By the above transformation, any SIMO OIC $\mathbf{Y}=X\mathbf{h}+\mathbf{Z}$ is equivalent to a SISO OIC $\widetilde{Y}=\sqrt{\trans{\mathbf{h}}\mathbf{h}}X+\widetilde{Z}$, as in \cite[Eq. (6)]{dytso2019MIMO}.
	\hfill\remarksymbol
\end{remark}

It is can be seen that the above reduction approach generally applies to any linear vector Gaussian channel with arbitrary input constraints. Nevertheless, in the following, we will present a useful property of the equivalent channel matrix $\tilde{\mathbb{H}}$ exclusively for the MIMO OWC. To this end, we first introduce the following definition.
\begin{definition}[Channel Graph]
	For any channel realization $\mathbb{H}$ (or equivalently $\tilde{\mathbb{H}}$), the \textit{channel graph} $\Gamma(\trans{\mathbb{H}}\mathbb{H})=G\left(\mathcal{V},\mathcal{E}\right)$ is the directed graph of the gram matrix $\trans{\mathbb{H}}\mathbb{H}$, where the vertices set $\mathcal{V}\triangleq \left\{ P_1,\cdots,P_{\nt}\right\}$ and the edge set 
	$\mathcal{E}\triangleq \left\{ (P_i,P_j):\trans{\mathbf{h}}_i\mathbf{h}_{j}\neq0 \right\}$.
\end{definition}

Without any loss, we may assume that the channel graph of our interest is \textit{strongly connected}\footnote{For any channel graph ${G}$ that is not strongly connected, we can find $\kappa $ connected components of the basis graph as $G_1=\left(\mathcal{V}_{\mathcal{I}_1}, \mathcal{E}_{1}\right), \cdots, {G}_{\kappa}=\left(\mathcal{V}_{\mathcal{I_{\kappa}}}, \mathcal{E}_{\kappa}\right)$. One can construct an orthogonal matrix $\mathbb{U}$ by subsequently arranging orthogonal basis of $\mathbb{H}_{\mathcal{I}_1}, \cdots, \mathbb{H}_{\mathcal{I}_{\kappa}}$. Then by left multiplying $\trans{\mathbb{U}}$ the channel matrix can be converted into a block diagonal one, and the orginal channel can be regarded as $\kappa$ parallel MIMO channels, whose channel graphs are strongly connected.}, i.e., there is no non-empty proper subset $\mathcal{U}\subsetneqq \mathcal{I}$ such that $
\trans{\mathbb{H}}_{ \mathcal{U}^{\text{c}} }\mathbb{H}_{\mathcal{U}}=\bm{0}
$.

Note that entries of the equivalent channel matrix $\tilde{\mathbb{H}}$ may take values over the field of real numbers. Despite this fact, the following theorem indicates that all entries of the first channel vector $\trans{\mathbf{v}_1}$ for the strongest eigen-subchannel that corresponds to the maximum singular value $\sigma_1$ has the same sign.
\begin{theorem}[Positivity of Strongest Eigen-subchannel]\label{thm:rankone}
	If the channel graph is strongly connected, then the first column vector of $\mathbb{V}_1$, i.e., $\mathbf{v}_1$, is either all positive or all negative, and $\sigma_1>\sigma_i$ for $i\in \{2,\cdots,r\}$. 
\end{theorem}
\begin{IEEEproof}
Since a nonnegative square matrix is irreducible if and only if its directed graph is strongly connected~\cite{Horn2012}. Thus, $\trans{\mathbb{H}} \mathbb{H}$ is irreducible. Note that $\mathbf{v}_1$ is the eigenvector of $\trans{\mathbb{H}}\mathbb{H}$ with respect to the maximum nonzero eigenvalue $\sigma_1^2$, i.e., the spectral radius of  $\trans{\mathbb{H}} \mathbb{H}$. Then we can conclude the theorem by using Perron-Frobenius theorem~\cite{Horn2012}.
\end{IEEEproof}

\begin{example}
	Consider a $4\times 4$ MIMO OWC with the following channel matrix 
			\begin{IEEEeqnarray}{rCl}
					\mathbb{H}
					&=&
					\begin{bmatrix}	
							0.451 & 	0.603 & 	1.310 & 	3.963 \\
							2.065 & 	1.504 & 	0.530 & 	0.245 \\ 
							5.401 & 	2.922 & 	1.877 & 	1.152 \\ 
							0.680 & 	2.633 & 	0.924 & 	3.632 
						\end{bmatrix}	.
					\nonumber 
				\end{IEEEeqnarray}
	By the channel reduction \eqref{eq:equiv}, the equivalent channel matrix is 
	\begin{IEEEeqnarray}{rCl}
			\widetilde{\mathbb{H}}
			=
			\begin{bmatrix}	
					5.096  &	4.086  &	2.473  &	4.131\\
					2.812  &	0.241  &	-0.136  &	-3.626\\
					-0.473  &	1.160  &	-0.488  &	-0.271\\
					-0.015  &	0.007  &	0.0391  &	-0.012 
				\end{bmatrix},
		\end{IEEEeqnarray}
It can be seen that the entries of the first row vector are all positive numbers, which verifies Theorem \ref{thm:rankone}. \hfill \remarksymbol
\end{example}


Without loss of generality, we may assume that $\mathbf{v}_1$ is positive throughout the paper. Then the first component of $\tilde{\mathbf{Y}}$ is given as $\tilde{{Y}}_1=\sigma_1 \trans{\mathbf{v}_1}\mathbf{X}+\tilde{Z}_1$, i.e., a MISO OIC with a nonnegative channel gain vector $\sigma_1 \trans{\mathbf{v}_1}$. Thus, it is seen that each MIMO OIC with a strongly connected channel graph is dominated (in the sense of singular values) by a MISO OIC. Implicitly, we also conclude that \textit{any rank-one MIMO OIC is equivalent to a MISO OIC (rather than a general MISO Gaussian channel), regardless of the form of input constraints}.
%

\subsection{Problem Formulation}

With assumptions of perfect channel state information available at both the transmitter side and the receiver side, in this paper we mainly focus on the capacity of the MIMO optical intensity channel~(\ref{eqn:model1}) under equal-cost constraints \eqref{eqn:ecc}. The single-letter capacity expression of such a memoryless channel is given by
\begin{equation}
	\mathsf{C}_{\mathsf{E}}\left(\widetilde{\mathbb{H}},\bm{\alpha},\sigma \right)
	=\sup_{{\mathscr{P}}_{\mathbf{X}} \textnormal{ satisfying }  \eqref{eqn:ecc}} \II\left(\mathbf{X};\widetilde{\mathbf{Y}}\right) ,
\end{equation}
where $\mathscr{P}_{\mathbf{X}}$ denotes the probability measure of any feasible input.

\section{Capacity Results for General MIMO EC-OIC}\label{sec:3}
We begin with a useful property that will be repeatedly utilized in the remainder of the paper. Among all continuous random variables $X$ satisfying $\expec{X}=\alpha$ and $0\le X \le 1$, the maximum-entropy distribution is the truncated exponential with the expectation $\alpha$ \cite{lapidothmoserwigger09_7}, and the maximized differential entropy is given by
	\begin{flalign}
		\mathfrak{h}_{\text{TE}}(\alpha)=-\left( 1 - \alpha\right)\mu_{\alpha}-\ln
		\left(1-\alpha \mu_{\alpha} \right)
	\end{flalign}
	in nats, where the nonnegative number $\mu_{\alpha}$ is the unique solution to the following equation:
	\begin{flalign}
		\alpha=\frac{1}{\mu_{\alpha}}-\frac{1}{\exp\left(\mu_{\alpha}\right)-1 }.
	\end{flalign}

\subsection{Existing Result on Low-SNR Asymptotics}
In this subsection, we revisit the existing result on the low-SNR capacity slope of the general MIMO EC-OIC \cite{chaabanrezkialouini18_2}. For any real-valued linear vector Gaussian channel $\mathbf{Y}=\mathbb{H}\mathbf{X}+{\mathbf{Z}}$, we define the equivalent input
$\mathbf{S}\triangleq\mathbb{H}\mathbf{X}=\trans{\left(S_1,\cdots,S_{\nr}\right)}$, whose covariance matrix is given by
\begin{equation}
	\mathbb{K}_{\mathbf{S}\mathbf{S}}\triangleq \mathbb{E}\left[\left(\mathbb{H}\mathbf{X}-\mathbb{H}\bm{\alpha}\right)\trans{\left(\mathbb{H}\mathbf{X}-\mathbb{H}\bm{\alpha}\right)}\right] 
\end{equation}
with the trace
\begin{flalign}
	\mathrm{tr}\left(\mathbb{K}_{\mathbf{S}\mathbf{S}}\right)&=\mathbb{E}\left[\mathrm{tr}\left(\left(\mathbb{H}\mathbf{X}-\mathbb{H}\bm{\alpha}\right)\trans{\left(\mathbb{H}\mathbf{X}-\mathbb{H}\bm{\alpha}\right)}\right)\right]\label{eq:37}\\
	&=\mathbb{E}\left[\mathrm{tr}\left(\trans{\left(\mathbb{H}\mathbf{X}-\mathbb{H}\bm{\alpha}\right)}\left(\mathbb{H}\mathbf{X}-\mathbb{H}\bm{\alpha}\right)\right)\right]\label{eq:39}\\
	&=\sum_{j=1}^{\nr} \mathsf{var}\left(S_j\right), 
\end{flalign}
where~\eqref{eq:39} follows from the cyclic invariance of the trace of matrix product, and $\mathsf{var}(\cdot)$ denotes the variance of some random variable.


It is well-known that the capacity slope of the real-valued linear vector Gaussian channel at zero SNR is half of the maximum trace of the covariance matrix over all feasible equivalent inputs $\widetilde{\mathbb{H}}\mathbf{X}$ \cite{prelovverdu04_1,limoserwangwigger20_1}. In \cite{chaabanrezkialouini18_2}, the authors have shown that the maximum trace for the MIMO OIC with equal-cost constraints is achieved by \textit{the maximally correlated multivariate binary distribution} (see \cite[Def. $1$]{chaabanrezkialouini18_2} for a detailed description). Also, in~\cite{chen2022MISO}, the same result is rediscovered for the MISO case via convex ordering.

%

 In the following, we reformulate the result on the maximum trace for the MIMO EC-OIC in \cite{chaabanrezkialouini18_2}, as well as an alternative proof.
	
	\begin{theorem}[\!\cite{chaabanrezkialouini18_2}]\label{lowsnr:ec}
		Over all inputs $\mathbf{X}$ with $\supp\, \mathbf{X} \subseteq [0,1]^{\nt}$ and the expectation $\bm{\alpha}$, the maximum trace of $\mathbb{K}_{\mathbf{S}\mathbf{S}}$ is given by
		\begin{equation}\label{eq:40}
			\setlength{\abovedisplayskip}{8pt}
			\const{V}_{\textnormal{max}}^{\textnormal{E}}\left(\mathbb{H},\bm{\alpha}\right)=\sum_{i=1}^{\nt}\sum_{j=1}^{\nt} g_{ij} \left(\min\{\alpha_i,\alpha_j\} -\alpha_i\alpha_j\right),	
		\end{equation}
		where $g_{ij}$ denotes the entry in the $i$-th row and $j$-th column of the Gram matrix $\mathbb{G}=\trans{\mathbb{H}}\mathbb{H}$. 
		\begin{IEEEproof}
			Assume that the transmitters are ordered such that $\alpha_1\ge \alpha_2 \ge \cdots \ge \alpha_{\nt}$. Based on the result in~\cite{chen2022MISO}, for any nonnegative channel matrix $\mathbb{H}\in \mathcal{M}_{\nr \times \nt}$, $\nr$ variances $\mathsf{var}\left(S_j\right)$ for $j\in [\nr]$ are maximized by a same maximally correlated $\nt$-variate binary distribution $\overbar{\mathbf{X}}_{\bm{\alpha}}$ with $\nt+1$ mass points $\bm{0}$, $\mathbf{e}_1$, $\cdots$, $\sum_{i=1}^{\nt-1}\mathbf{e}_{i}$, and $\sum_{i=1}^{\nt}\mathbf{e}_{i}$ and corresponding probabilities $ 1-\alpha_1$, $\alpha_1-\alpha_2$, $\cdots$, $\alpha_{\nt-1}-\alpha_{\nt}$, and $\alpha_{\nt}$, where $\mathbf{e}_{i}$ denotes the $i$-th column vectors of the $\nt\times \nt$ identity matrix. Hence, the sum of those $\nr$ variances, i.e., $\mathrm{tr}\left(\mathbb{K}_{\mathbf{S}\mathbf{S}}\right)$, is also maximized by $\overbar{\mathbf{X}}_{\bm{\alpha}}$. Substituting the probability mass function of $\overbar{\mathbf{X}}_{\bm{\alpha}}$ into \eqref{eq:37}, we get \eqref{eq:40}.
		\end{IEEEproof}
	\end{theorem}
	
	Then the following capacity upper bound and asymptotics follows from a standard argument that Gaussian maximizes the differential entropy.
	\begin{corollary}[\cite{chaabanrezkialouini18_2}]
		\label{ecmc:max-var-bnd}
		The capacity of the MIMO optical intensity channel~(\ref{eqn:model1}) under equal-cost constraints \eqref{eqn:ecc} is upper-bounded as
		\begin{flalign}\label{eq:max-var-bnd}
		\mathsf{C}_{\mathsf{E}}\left(\widetilde{\mathbb{H}},\bm{\alpha},\sigma \right)
		\le \frac{r}{2}\log \left(1+\frac{\const{V}_{\textnormal{max}}^{\textnormal{E}}\left(\mathbb{H},\bm{\alpha}\right)}{r\sigma^2}\right),
		\end{flalign} 
		and the low-SNR capacity slope is given by
			\begin{flalign}
			\lim_{\sigma\to +\infty}   \sigma^2
			\mathsf{C}_{\mathsf{E}}\left(\widetilde{\mathbb{H}},\bm{\alpha},\sigma \right)
			= \frac{\const{V}_{\textnormal{max}}^{\textnormal{E}}\left(\mathbb{H},\bm{\alpha}\right)}{2}.
		\end{flalign}
		\begin{IEEEproof}
			Omitted.
		\end{IEEEproof}
\end{corollary}


\subsection{Achievable Rate for Linear Precoding}\label{subsec:linear_precoding_ecoic}
In the remainder of this section, we turn to achievable rates of our considered channel. Unlike the existing model relying on a maximum-entropy argument, characterizing the maximum differential entropy of the equivalent input $\tilde{\mathbb{H}}\mathbf{X}$ under per-antenna intensity constraints is generally difficult when $\tilde{\mathbb{H}}$ has a full row rank, i.e., underdetermined\footnote{In~\cite{mosermylonakiswangwigger17_1,chaabanrezkialouini18_1}, the MIMO OIC of full column rank has been investigated. In this special case, by using the channel transformation in Sec. \ref{sec:reduction}, the transformed channel output is the summation of the AWGN and the image of the channel input $\mathbf{X}$ under an invertible linear transformation $\widetilde{\mathbb{H}}$, which implies that the high-SNR asymptotic capacity is only determined by the maximum differential entropy of $\mathbf{X}$ even in the regimes of no CSIT. However, such the statement may no longer hold for the rank-deficient case.}.

In this subsection, we propose a simple linear precoding scheme, based on which the original full-row-rank channel is simplified to be invertible. The main idea is letting the channel input be the linear combination of $r$ channel inputs that correspond to $r$ transmitters with linearly independent channel column vectors.

We first choose a size-$r$ index set $\mathcal{I}$ such that $\det\left(\mathbb{H}_{\mathcal{I}}\right)\neq 0$. Without loss of generality, by an appropriate re-ordering of all transmitters, we may let $\mathcal{I}=[r]$. For notational simplicity, we define $\bm{\beta}\triangleq \bm{\alpha}-\frac{1}{2}\bm{1}_{\nt}$.

Let $\mathbf{X}_{\mathcal{I}}$ denote some feasible input of the transmitters indexed by $\mathcal{I}$ and 
 input of others transmitters be some linear combination of $\mathbf{X}_{\mathcal{I}}$ as follows:
\begin{flalign}
	X_{j+r}-\frac{1}{2}=\trans{\mathbf{b}_j}\left(\mathbf{X}_{\mathcal{I}}-\frac{1}{2}\mathbf{1}\right)+c_j,~j \in [\nt-r].
\end{flalign}
It is clear that $ -\frac{1}{2}\left\| \mathbf{b}_j \right\|_1   +c_j \le  X_{j+r} -\frac{1}{2} \le \frac{1}{2}\left\| \mathbf{b}_j \right\|_1 +c_j $ due to the amplitude constraint \eqref{eqn:ecc1}. Hence, to satisfy such an amplitude constraint, it is sufficient that
\begin{subequations}\label{eq:support_linear}
	\begin{align}
		-\frac{1}{2}\left\| \mathbf{b}_j \right\|_1  +c_j &\ge -\frac{1}{2}, \\
		\frac{1}{2}\left\| \mathbf{b}_j \right\|_1 +c_j &\le \frac{1}{2} .
	\end{align}	
\end{subequations}
For the EC-OIC, it is required to satisfy the average-intensity equality as
\begin{flalign}\label{eq:equality_linear}
	\trans{\mathbf{b}_j}\bm{\beta}_{\mathcal{I}}+c_j&=\beta_{j+r},~j \in [\nt-r].
\end{flalign}

Note that \eqref{eq:support_linear} immediately leads to $\left\| \mathbf{b}_j \right\|_1\le 1$, and the existence of $c_j$ is guaranteed by $\left\| c_j \right\|\le \frac{1}{2}\left( 1-\left\| \mathbf{b}_j \right\|_1 \right)$. Thus, the dual intensity constraint of the EC-OIC is satisfied if and only if
\begin{subequations}\label{eq:17}
	\begin{align}
		\left\| \mathbf{b}_j \right\|_1    &\le 1, \\
		\left|	\trans{\mathbf{b}_j}\bm{\beta}_{\mathcal{I}}-\beta_{j+r} \right|  & \le  \frac{1}{2}-\frac{1}{2}\left\| \mathbf{b}_j \right\|_1,
	\end{align}	
\end{subequations}
i.e., the original inequality involving the weighting vector $\mathbf{b}_j$ and the bias constant $c_j$ is equivalently converted into one involving only $\mathbf{b}_j$.

Define the precoding matrix as $\mathbb{B}\triangleq \left[\mathbf{b}_1,\cdots,\mathbf{b}_{\nt-r} \right]$. Then, by using such a linear precoding strategy, the channel output is given by
\begin{flalign}\label{eq:linear_transmission}
	\mathbf{Y}=\left(\mathbb{H}_{\mathcal{I}}+\mathbb{H}_{\mathcal{I}^{c}}\trans{\mathbb{B}} \right)\mathbf{X}_{\mathcal{I}}+ \mathbf{Z}.
\end{flalign}
From the viewpoint of maximizing the differential entropy of the equivalent input $\left(\mathbb{H}_{\mathcal{I}}+\mathbb{H}_{\mathcal{I}^{c}}\trans{\mathbb{B}}  \right)\mathbf{X}_{\mathcal{I}}$, we are equivalently concerned with the following optimization problem 
\begin{equation}\label{prob:max_det}
	\begin{aligned}
		&\textnormal{max} ~&&\quad\quad~ 
		\left| \det \left( \mathbb{H}_{\mathcal{I}}+\mathbb{H}_{\mathcal{I}^{c}}\trans{\mathbb{B}}  \right) \right|
		\\
		&\textnormal{s.t.}~ &&~~\quad\quad\quad \left\| \mathbf{b}_j  \right\|_{1} \le 1,\\
		& &&  \left|	\trans{\mathbf{b}_j}\bm{\beta}_{\mathcal{I}}-\beta_{j+r} \right|  \le  \frac{1}{2}-\frac{1}{2}\left\| \mathbf{b}_j \right\|_1,~j\in[\nt-r].
	\end{aligned}
\end{equation}
Generally, the square matrix $\mathbb{H}_{\mathcal{I}}+\mathbb{H}_{\mathcal{I}^{c}}\mathbb{B} $ is not positive semi-definite and hence the above problem is not convex.

Owing to the separability of constraints on the precoding matrix $\mathbb{B}$, we propose an iterative optimization algorithm in what follows.

For any invertible matrix and any pair of real vectors $\mathbf{u}$ and $\mathbf{v}$, the following equality holds
$$
\det\left( \mathbb{A}  +  \mathbf{u}\trans{\mathbf{v}} \right)
= \left( 1+ \trans{\mathbf{v}} \mathbb{A}^{-1} \mathbf{u}  \right)\det\left( \mathbb{A}\right)
$$
due to the matrix determinant lemma \cite{Horn2012}. Thus, for any invertible matrix $\mathbb{A}$, if maximize $\left|\det \left( \mathbb{A}+\mathbf{h}_{j+r}\trans{\mathbf{b}_j} \right)\right|$ under the constraint on the precoding vector $\mathbf{b}_j$ as in \eqref{prob:max_det}, we only need to solve the following problem
\begin{subequations}\label{prob:iter}
	\begin{align}
		\textnormal{max} ~~&\quad\quad 
		\left| 1+ \trans{\mathbf{b}_j}  \mathbb{A}^{-1} \mathbf{h}_{j+r} \right|
		\\
		\textnormal{s.t.} &\quad\quad~~\quad \left\| \mathbf{b}_j  \right\|_{1} \le 1, \label{eq:20b}\\
		  &  \left|	\trans{\mathbf{b}_j}\bm{\beta}_{\mathcal{I}}-\beta_{j+r} \right|  \le  \frac{1}{2}-\frac{1}{2}\left\| \mathbf{b}_j \right\|_1, \label{eq:20c}
	\end{align}
\end{subequations}
whose solution is denoted by $\mathbf{b}_{j}^{\star}$.

For simplicity, we introduce some notations in what follows. Let $\mathbf{w} \triangleq \mathbb{A}^{-1} \mathbf{h}_{j+r}$ and $\mathcal{K}$ be the set consisting of all indices $k$ such that $w_k\ge 0$. Let $K$ be the cardinality of $\mathcal{K}$ and $\mathcal{K}^{\text{c}}= [r]\setminus \mathcal{K}$. Then we define constant vectors $\mathbf{w}_{1}\triangleq\mathbf{w}_{\mathcal{K}}$, $\mathbf{w}_{2}\triangleq -\mathbf{w}_{\mathcal{K}^{\text{c}}}$, $\bm{\alpha}_1=\bm{\alpha}_{\mathcal{K}}$, and $\bm{\alpha}_2=\bm{\alpha}_{\mathcal{K}^{\text{c}}}$, and two variables $\bm{\mu}_{1}\in \mathbb{R}_{+}^{K}$ and $\bm{\mu}_2 \in \mathbb{R}_{+}^{r-K}$. The optimal solution $\mathbf{b}_{j}^{\star}$ to the $\ell_1$-norm maximization problem \eqref{prob:iter} can be efficiently solved by using the following proposition.

\begin{proposition}\label{prop:linear_ec}
	For $\mathsf{t}\in \left\{1,2\right\}$, denote the solution and the optimal value of the following linear programming problem
	\begin{subequations}\label{prob:iter2}
		\begin{align}
			\textnormal{max} ~~&\quad\quad 
			\trans{\mathbf{w}}_{\mathsf{t}}\bm{\mu}_{\mathsf{t}}
			\\
			\textnormal{s.t.} ~~
			& \quad\quad \bm{\mu}_{\mathsf{t}}\succcurlyeq \bm{0} \label{eq:21b}\\
			&\quad\quad \trans{\bm{\alpha}_{\mathsf{t}}}\bm{\mu}_{\mathsf{t}}\le \alpha_{j+r},  \label{eq:21d}\\
			&\quad\quad \left(\trans{\bm{1}}-\trans{\bm{\alpha}_{\mathsf{t}}}\right)\bm{\mu}_{\mathsf{t}}\le 1-\alpha_{j+r}, \label{eq:21e}
		\end{align}
	\end{subequations}
	by $\bm{\mu}_{\mathsf{t}}^{\star}$ and $\const{L}_{\mathsf{t}}^{\star}$. If $1+\trans{\mathbf{w}}_{1}\bm{\mu}_{1}\ge \left|1-\trans{\mathbf{w}}_{2}\bm{\mu}_{2}\right|$, then $\mathbf{b}_{j,\mathcal{K}}^{\star}=\bm{\mu}_1$ and $\mathbf{b}_{j,\mathcal{K}^{\text{c}}}^{\star}=\bm{0}$; otherwise, $\mathbf{b}_{j,\mathcal{K}^{\text{c}}}^{\star}=\bm{\mu}_2$ and $\mathbf{b}_{j,\mathcal{K}}^{\star}=\bm{0}$.
\end{proposition}

\begin{IEEEproof}
See Appendix \ref{app:linear_ec}.
\end{IEEEproof}

Then we rewrite the objective function in \eqref{prob:max_det} as
\begin{flalign}
	\det \left( \mathbb{H}_{\mathcal{I}}+\mathbb{H}_{\mathcal{I}^{c}}\trans{\mathbb{B}}  \right)  
	=  \det \left( \mathbb{H}_{\mathcal{I}}+
	\sum_{k=1}^{\nt-r} \mathbf{h}_{k+r}\trans{\mathbf{b}_k}
	\right),
\end{flalign} 
and a suboptimal solution $\mathbb{B}_{\mathsf{E}}^{\dag}$ to \eqref{prob:max_det} can be obtained by iteratively letting $\mathbb{A}=\mathbb{H}_{\mathcal{I}}+
\sum_{k\in[\nt-r]\setminus\left\{ j\right\}} \mathbf{h}_{k+r}\trans{\mathbf{b}_k}$ and individually optimizing $\mathbf{b}_j$ for all $j\in [\nt-r]$. For clarity, we summarize the above iterative optimization method in Algorithm 1, for which the standard toolbox, such as CVX, can be readily used to efficiently solve the involving linear programming problems.

Let $\const{D}_{\mathsf{E}}=\left| \det \left( \mathbb{H}_{\mathcal{I}}+\mathbb{H}_{\mathcal{I}^{c}}\trans{\left(\mathbb{B}^{\dag}_{\mathsf{E}}\right)}  \right)\right| $. Then an achievable rate of the above linear transmitter architecture is given by the following proposition.
\begin{proposition}\label{prop:linear_precoding}
	The achievable rate of the MIMO optical intensity channel \eqref{eq:linear_transmission} with linear precoding and under equal-cost constraints \eqref{eqn:ecc} is lower-bounded as 
	\begin{flalign}\label{eq:linear_precoding}
		\mathsf{C}_{\mathsf{E}}\left(\widetilde{\mathbb{H}},\bm{\alpha},\sigma\right)
		\ge
		\frac{ r }{2}\log\left(1+
		\frac{\sqrt[r]{\const{D}_{\mathsf{E}}^2}\exp\left( 2\left(\sum_{i=1}^{r}\mathfrak{h}_{\text{TE}}(\alpha_i)\right)/r \right)}{2\pi e \sigma^2}\right).
	\end{flalign}
\end{proposition}
\begin{IEEEproof}
	 Note that the differential entropy of the equivalent input $\mathbf{S}=\left(\mathbb{H}_{\mathcal{I}}+\mathbb{H}_{\mathcal{I}^{c}}\trans{\left(\mathbb{B}^{\dag}_{\mathsf{E}}\right)}  \right)\mathbf{X}_{\mathcal{I}}$ is $\log\const{D}_{\textsf{a}}+\sum_{i=1}^{r}\hh(X_i)$, which is maximized by letting $\hh(X_i)=\mathfrak{h}_{\text{TE}}(\alpha_i)$ for all $i\in[r]$ \cite{coverthomas06_1}. The proposition is concluded by using the entropy power inequality 
	 \begin{flalign}\label{eq:epi}
	 	\exp\left(  \frac{2}{r} \hh\left(\mathbf{S}+\mathbf{N}\right) \right)\ge 	\exp\left(  \frac{2}{r} \hh\left(\mathbf{S}\right)  \right)+	\exp\left(  \frac{2}{r} \hh\left(\mathbf{N}\right)  \right),
	 \end{flalign}
	 for any two independent $r$-dimensional random vectors $\mathbf{S}$ and $\mathbf{N}$ with non-degenerate differential entropies.
\end{IEEEproof}

\begin{algorithm}\label{alg:CIO}
	\caption{Circularly Iterative Optimization for \eqref{prob:max_det}}
	\begin{algorithmic}[1]	
		\Require
		$\bm{\alpha}$, $\mathbb{H}$, $n_{\text{cycle}}$
		\Ensure Precoder matrix $\mathbb{B}$ 
		\kern2pt\hrule\kern2pt
		\State $\mathbb{B}=\bm{0}_{r\times (\nt-r)}$;
		\For{$p=1:1:n_{\text{cycle}}$}
		\For{$j=1:1:\nt-r$}
		\State $\mathbb{A}=\mathbb{H}_{\mathcal{I}}+\mathbb{H}_{\mathcal{I}^{c}}\mathbb{B} -\mathbf{h}_{j+r}\trans{\mathbf{b}_j}$;
		\State $\mathbf{b}=\bm{0}_r$;
		\State $\mathbf{w} = \mathbb{A}^{-1} \mathbf{h}_{j+r}$;
		\State $\mathcal{K}=\left\{ k: w_k\ge 0\right\}$;
		\State $\mathcal{K}^{\text{c}}= [r]\setminus \mathcal{K}$;
		\State $\mathbf{w}_{1}=\mathbf{w}_{\mathcal{K}}$;
		\State $\mathbf{w}_{2}= -\mathbf{w}_{\mathcal{K}^{\text{c}}}$;
		\State $\bm{\alpha}_1=\bm{\alpha}_{\mathcal{K}}$;
		\State $\bm{\alpha}_2=\bm{\alpha}_{\mathcal{K}^{\text{c}}}$;
		\State solve linear programming problems \eqref{prob:iter2} by CVX.
		\If{$1+\trans{\mathbf{w}}_{1}\bm{\mu}_{1}^{\star}\ge \left|1-\trans{\mathbf{w}}_{2}\bm{\mu}_{2}^{\star}\right|$}
		\State $\mathbf{b}_{\mathcal{K}}=\bm{\mu}_{1}^{\star}$;
		\Else
		\State $\mathbf{b}_{\mathcal{K}^{\text{c}}}=\bm{\mu}_{2}^{\star}$;
		\EndIf
		\State $\mathbf{b}_{j}=\mathbf{b}$;
		\EndFor
		\EndFor
	\end{algorithmic}
\end{algorithm}
\subsection{Achievable Rate Based on Generalized EPI}\label{sec:gepi_ec}

In this subsection, another achievable rate is derived by the generalized entropy-power inequality in \cite{zamir1993gepi}.
Assume that all $X_i$ are independent, then the differential entropy of the equivalent input satisfies
\begin{flalign}
	\hh(\mathbb{H}\mathbf{X})&\ge \hh(\mathbb{H}\mathbf{N}) \label{eq:gepi} \\
	&=\frac{1}{2}\log\left(\left| \det(\mathbb{H}\mathbb{D}\trans{\mathbb{H}})  \right|\right)\\
	&=\log\left| \det \left( \sum_{i=1}^{\nt}d_i \cdot\left( \mathbf{h}_i\trans{\mathbf{h}}_i\right) \right)\right| \label{eq:31}
\end{flalign}
where \eqref{eq:gepi} follows from the generalized EPI, the $\nt$-dimensional zero-mean Gaussian vector $\mathbf{N}$ has independent components $N_i\sim\mathcal{N}(0,\exp(2\hh(X_i))/(2\pi e))$,  the diagonal matrix $\mathbb{D}=\diag{\exp(2\hh(X_1)) ,\cdots,\exp(2\hh(X_{\nt})) }$, and $d_i=\exp(2\hh(X_i))$ for all $i\in[\nt]$.

Define $\mathbf{d}_{\max} \triangleq \left(\exp(2\mathfrak{h}_{\text{TE}}(\alpha_1)) ,\cdots,\exp(2\mathfrak{h}_{\text{TE}}(\alpha_{\nt}))\right)$. It is straightforward that $\bm{0} \prec \mathbf{d} \preccurlyeq  \mathbf{d}_{\max} $. The following lemma presents the condition when the right-hand side of \eqref{eq:31} is maximized.

\begin{lemma}\label{lemma:1}
	For $\mathbf{d}$ satisfying $ \bm{0} \prec  \mathbf{d}\preccurlyeq \mathbf{d}_{\max}$, the logarithmic determinant $\log\left|\det \left( \sum_{i=1}^{\nt}d_i \cdot\left( \mathbf{h}_i\trans{\mathbf{h}}_i\right) \right)\right|$ is maximized when $\mathbf{d}=\mathbf{d}_{\max}$.
\end{lemma}
\begin{IEEEproof}
Note that $
	\mathbb{H}\mathbb{D}\trans{\mathbb{H}}=\sum_{i=1}^{\nt}d_i \cdot\left( \mathbf{h}_i\trans{\mathbf{h}}_i\right)$
	is positive semi-definite for any $d_i\ge 0$. Thus, the inequality $\det \left( \sum_{i=1}^{\nt}d_i \cdot\left( \mathbf{h}_i\trans{\mathbf{h}}_i\right) \right)\ge 0$ always holds.
	Let $\mathbb{X}=\sum_{i=2}^{\nt}d_i \cdot\left( \mathbf{h}_i\trans{\mathbf{h}}_i\right)$ and $\mathbb{Y}=\mathbf{h}_1\trans{\mathbf{h}}_1$. It is well known that the derivative of the logarithmic determinant is given by
	$\frac{\dd}{\dd t} \log\det \left( \mathbb{X}+t\mathbb{Y} \right)= \trace{ \left( \mathbb{X}+t\mathbb{Y} \right)^{-1}\mathbb{Y}}$. Due to $\mathbf{d} \succ \bm{0}$, three square matrices $\mathbb{X}+t\mathbb{Y}$, $\left( \mathbb{X}+t\mathbb{Y} \right)^{-1}$ and $\mathbb{Y}$ are all positive semi-definite for any $t>0$. By using the trace inequality in \cite[Theorem 4.3.53]{Horn2012} for two positive semi-definite matrices, we have $\trace{ \left( \mathbb{X}+t\mathbb{Y} \right)^{-1}\mathbb{Y}}\ge 0$, which indicates $\log\det \left( \mathbb{X}+t\mathbb{Y} \right)$ is maximized when $t=\exp(2\mathfrak{h}_{\text{TE}}(\alpha_1))$ for any feasible choice of $\mathbb{X}$. The lemma is concluded by similarly applying the above technique to $d_2$, $\cdots$, and $d_{\nt}$. 
\end{IEEEproof}

Using the same proof technique for Proposition \ref{prop:linear_precoding} and Lemma \ref{lemma:1}, we obtain the following capacity lower bound.
\begin{proposition}[GEPI-Based Lower Bound]\label{prop:gepi_ec}
		The capacity of the MIMO optical intensity channel~(\ref{eqn:model1}) under equal-cost constraints \eqref{eqn:ecc} is lower-bounded as
	\begin{flalign}
		\mathsf{C}_{\mathsf{E}}\left(\widetilde{\mathbb{H}},\bm{\alpha},\sigma\right)
		\ge
		\frac{r}{2}\log\left(1+\frac{ \sqrt[r]{\det(\mathbb{H}\cdot\diag{\mathbf{d}_{\text{max}}} \cdot\trans{\mathbb{H}})} }{2\pi e \sigma^2}\right).
	\end{flalign}
	\begin{IEEEproof}
		Omitted.
	\end{IEEEproof}
\end{proposition}

\subsection{QR Receiver With Successive Interference Cancellation}\label{sec:qr-sic}
In \cite{chen2022MISO}, the authors have addressed the capacity problem for the MISO OIC under per-antenna intensity constraints. Combining with the idea of QR decomposition and successive interference cancellation, we may borrow capacity results on the MISO OIC from \cite{chen2022MISO} to develop a simple transmission scheme that achieves full multiplexing gains.

Denote the QR decomposition of the channel matrix $ \widetilde{\mathbb{H}}$ as $
\widetilde{ \mathbb{H}} = \mathbb{Q} \mathbb{P}
$, where $\mathbb{Q}$ is an $r \times r$ orthogonal matrix and $\mathbb{P}=[p_{ij}]\in \mathbb{P}^{r \times \nt}$ is an upper triangular matrix with nonnegative diagonal entries. The QR receiver outputs
\begin{flalign}\label{eq:qr_receiver}	\tilde{\mathbf{Y}}=\trans{\mathbb{Q}}\mathbf{Y}=\mathbb{P}\mathbf{X}+\tilde{\mathbf{Z}},
\end{flalign}
which can be further regarded as $r$ vector subchannels as follows 
\begin{flalign}
	\tilde{Y}_{i}=\sum_{k=i}^{\nt} p_{ik}X_i+\tilde{Z}_i, ~\text{for } i\in [r].
\end{flalign} 
Let $\mathbf{g}\triangleq\left(p_{\tau,\tau},\cdots,p_{\tau,\nt} \right)\in \mathbb{R}^{\nt-\tau+1}$, $\mathcal{U}=\left\{i\in [\nt]: p_{\tau,i}<0,~i\ge \tau \right\}$, and define the vector $\check{\mathbf{\alpha}}$ by $\check{\mathbf{\alpha}}_{\mathcal{U}}=\bm{1}-\mathbf{\alpha}_{\mathcal{U}}$ and $\check{\mathbf{\alpha}}_{\mathcal{U}^{\text{c}}}= \mathbf{\alpha}_{\mathcal{U}^{\text{c}}}$. It is noted that the new channel coefficients $p_{ik}$ take values on $\mathbb{R}$ rather than $\mathbb{R}_+$. For this reason, for negative channel coefficient $p_{\tau,k}$ ($k\ge \tau$) of the $\tau$-th subchannel, we flip its sign by rewritten as $X_k=1-\tilde{X}_k$. 

At the transmitter side, we let $(X_{r},\cdots,X_{\nt})$ jointly be the capacity-achieving input of the $r$-th subchannel, i.e., a MISO EC-OIC with channel gains $\left|\mathbf{g}\right|$ and average intensities $\breve{\bm{\alpha}}$, and the input $X_{i}$ (for all $i\in [r-1]$) independently be the capacity-achieving input of the scalar optical intensity channel $Y=X+Z$ with input constraints $X\in[0,1]$ and $\mathbb{E}[X]=\alpha_i$, respectively. At the receiver side, we first decode $(X_{r},\cdots,X_{\nt})$ from the $\tilde{Y}_r$, and then subsequently decode $X_{r-1}$, $\cdots$, and $X_{1}$ from the other subchannels $\tilde{Y}_{r-1}$, $\cdots$, and $\tilde{Y}_1$ via successive interference cancellation. The achievable rate of such a QR-SIC scheme is given in the following proposition. 

\begin{proposition}\label{prop:ec_oic_qr}
	For the MIMO optical intensity channel \eqref{eq:linear_transmission} under bounded-cost constraints \eqref{eqn:bcc}, the achievable rate of the QR-SIC scheme is given by
	\begin{flalign}\label{eq:ec_oic_qr}
\const{R}_{\text{QR}}= \mathsf{C}_{\mathsf{E}}\left(\trans{\left|\mathbf{g}\right|},\check{\bm{\alpha}},\sigma\right)+\sum_{i=1}^{r-1}
\mathsf{C}_{\mathsf{E}}\left(p_{i,i},\alpha_{i},\sigma \right),
	\end{flalign}
where the involved capacities $\mathsf{C}_{\mathsf{E}}\left(p_{i,i},\alpha_{i},\sigma \right)$ of the MISO EC-OIC can be evaluated by results in \cite{chen2022MISO}.
	\begin{IEEEproof}
		Omitted.
	\end{IEEEproof}
\end{proposition}

\vspace{-1em}

\section{Equivalent Capacity Expressions for Rank-($\nt-1$)} \label{sec:equiv}
Although capacity expressions of the general MIMO OIC under a total average-intensity constraint and peak-intensity constraints and the MISO EC-OIC have been given in \cite{limoserwangwigger20_1} and \cite{chen2022MISO}, respectively, a general solution to that of the MIMO OIC under per-antenna intensity constraints still seems to be intractable. Even for the MISO case \cite{chen2022MISO}, the analysis relies on several infrequently-used concepts, such as the stop-loss function, convex ordering and comonotonicity. Unfortunately, the technique applied in the MISO case can not be generalized to our considered MIMO case since the optimality of comonotonic inputs no longer holds, e.g., independent input is asymptotically capacity-achieving when $\widetilde{\mathbb{H}}$ is invertible. 

Despite of the above challenges, in this subsection we characterize the capacity expression for the MIMO EC-OIC in the special case $r=\nt-1$ from the perspective of convex geometry that distinguished from the idea of distribution decomposition in~\cite{chen2022MISO}. 

As shown in \cite{limoserwangwigger20_1}, the admissible region $\mathcal{R}\left({\widetilde{\mathbb{H}}}\right)$ of the equivalent input $\mathbf{S}=\widetilde{\mathbb{\mathbb{H}}}\mathbf{X}$ under amplitude constraints \eqref{eqn:ecc1} on $\mathbf{X}$ is the zonotope $$
\mathcal{R}\left({\widetilde{\mathbb{H}}}\right)\triangleq\left\{ \sum_{i=1}^{\nt} x_i\widetilde{\mathbf{h}}_i:x_i\in[0,1] \textnormal{ for all }i\in [\nt]\right\}.
$$
Note that the volume of $\mathcal{R}\left({\widetilde{\mathbb{H}}}\right)$ {\color{black}is equal} to the determinant of the full-rank square matrix $\widetilde{\mathbb{H}}\trans{\widetilde{\mathbb{H}}}$, which implies $\mathcal{R}\left({\widetilde{\mathbb{H}}}\right)$ is non-degenerate in $r$-space. Moreover, it can be shown that the zonotope $\mathcal{R}\left({\widetilde{\mathbb{H}}}\right)$ is compact in Euclidean $r$-space.
The following theorem reveals that any MIMO EC-OIC of rank $\nt-1$ is equivalent to a vector Gaussian channel under with a zonotope support constraint, an expectation constraint, and two piecewise linear cost constraints.

\begin{theorem} [Equivalent Capacity Expression]
	\label{thm:nt-1} 
	Let $v_{j,i}$ denote the entry that lies in the $j$-column and the $i$-row of the orthogonal matrix $\mathbb{V}$. 
	In the case of $r=\nt-1$, the capacity of the MIMO optical intensity channel~(\ref{eqn:model1}) under equal-cost constraints \eqref{eqn:ecc} can be expressed as 
	\begin{equation}\label{ecc:capacity}
		\mathsf{C}_{\textnormal{E}}\left(
		\widetilde{\mathbb{H}},\bm{\alpha},\sigma \right)
		=\sup_{\mathscr{P}_{ \mathbf{S}}} \II \left(\mathbf{S};\mathbf{S}+\mathbf{Z}\right),
	\end{equation}
	where the supremum is over all probability laws $\mathscr{P}_{ \mathbf{S}}$ on $\mathbf{S}$ with $\supp\, \mathbf{S} \subseteq \mathcal{R}\left(\widetilde{\mathbb{H}}\right)$ and 
	\begin{subequations}\label{eq:equiv_ecmc}
		\begin{align}
			&\mathbb{E}\left[\mathbf{S}\right]=\widetilde{\mathbb{H}}\bm{\alpha},\label{eq:equiv_ecmc1}\\
			&\mathbb{E} \left[ f_{\textnormal{min}}(\mathbf{S})\right]
			\le  \trans{\mathbf{v}}_{\nt}\bm{\alpha}, 
			\label{eq:equiv_ecmc2}\\
			&
			\mathbb{E} \left[ f_{\textnormal{min}}\left(\widetilde{\mathbb{H}}\bm{1}-\mathbf{S}\right)\right] \le \trans{\mathbf{v}}_{\nt}\left(\bm{1}-\bm{\alpha}\right), 
			\label{eq:equiv_ecmc3}
		\end{align}	
	\end{subequations}
	where $\widetilde{\mathbf{h}}_i$ denotes the $i$-th column vector of $\widetilde{\mathbb{H}}$, and
	\begin{equation}
		f_{\textnormal{min}}(\mathbf{s})=\max_{ {\color{black}i:\,} v_{\nt,i}\neq 0}\left\{ \frac{1-\textnormal{sign}(v_{\nt,i})}{2v_{\nt,i}} -\sum_{j=1}^{\nt-1} \frac{v_{j,i}s_j}{v_{\nt,i}\sigma_j} \right\}.
	\end{equation}
	\begin{IEEEproof}
		See Appendix~\ref{app:nt-1}.
	\end{IEEEproof}
\end{theorem}

As presented in Theorem \ref{thm:nt-1}, the derived capacity expression remove the linear effect  brought by the rank-deficient channel matrix and equivalently convert the original model into a vector Gaussian channel under some moment constraints. It is well-known that the differential entropy of a random vector under moment constraints is maximized by the Maxwell–Boltzmann distribution \cite{coverthomas06_1}. Hence, an achievable rate for the rank-$(\nt-1)$ MIMO EC-OIC can be obtained by convex programming for the maximum differential entropy and the EPI.

\begin{theorem}
	\label{thm:highsnr_ECMC}
	The capacity of the MIMO EC-OIC of rank $r=\nt-1$ is lower-bounded by
	\begin{equation}
		\mathsf{C}_{\textnormal{E}}\left(\widetilde{\mathbb{H}},\bm{\alpha},\sigma\right)
		\ge
		\frac{ \nt-1 }{2}\log\left(1+
		\frac{\exp\left( 2\gamma_{\textnormal{E}}/\left(\nt-1 \right)\right)}{2\pi e \sigma^2}\right),
	\end{equation}
	where 
	\begin{flalign} \label{prob:max_ent}
		\gamma_{ \textnormal{E}} = 	& \min_{\nu \in \mathbb{R}, \mathbf{u}\in\mathbb{R}^{r} \atop \lambda_1,\lambda_{2}\ge 0}
		\trans{\mathbf{u}}\widetilde{\mathbb{H}}\bm{\alpha}+\lambda_1 \trans{\mathbf{v}}_{\nt}\bm{\alpha}
		+\lambda_2  \trans{\mathbf{v}}_{\nt} \left(\bm{1}-\bm{\alpha}\right)-1-\nu
		\nonumber\\
		+& \int_{\mathcal{R}\left(\widetilde{\mathbb{H}}\right)}  \exp \left(\nu -\trans{\mathbf{u}}\mathbf{s}-
		\lambda_1 f_{\textnormal{min}}\left(\mathbf{s}\right)-\lambda_2 f_{\textnormal{min}}\left(\widetilde{\mathbb{H}}\bm{1}-\mathbf{s}\right)
		\right) \dd \mathbf{s}.
	\end{flalign} 
	is the maximum differential entropy of the equivalent input $\mathbf{S}$ for the MIMO EC-OIC.
\end{theorem}
\begin{IEEEproof}
Along the same line of the proof \cite[Thm. 12.1.1]{coverthomas06_1}, the maximum differential entropy of distributions under a give support constraint and multiple moment constraints with measurable cost functions can be formulated as a variation problem, and then, is converted into the above convex problem by the Lagrangian dual.
We conclude the proof by applying the EPI \eqref{eq:epi} to lower-bound the differential entropy of $\mathbf{Y}$.
\end{IEEEproof}


\begin{remark} [Signaling for an EC-OIC of rank $\nt-1$] Given an equivalent input $\mathbf{S}$ satisfying~\eqref{eq:equiv_ecmc} and $\mathcal{S}=\supp\, \mathbf{S} \subseteq \mathcal{R}\left(\widetilde{\mathbb{H}}\right)$, the transmitters are only interested in the feasible input $\mathbf{X}$ such that $\widetilde{\mathbb{H}}\mathbf{X}=\mathbf{S}$, rather than $\mathbf{S}$. Herein, a very simple method of mapping $\mathbf{S}$ onto the desired $\mathbf{X}$ is demonstrated as follows. 
	\begin{itemize}
		\item \textit{Step 1:} Find the unique real number $\tau\in [0,1]$ satisfying
		$
			\tau \mathbb{E} \left[ f_{\textnormal{min}}(\mathbf{S})\right] + \left(1-\tau\right)\left( \trans{\bm{1}}\mathbf{v}_{\nt}-\mathbb{E} \left[ f_{\textnormal{min}}\left(\widetilde{\mathbb{H}}\bm{1}-\mathbf{S}\right)\right] \right) =\trans{\mathbf{v}}_{\nt}\bm{\alpha}.
		$
		\item \textit{Step 2:} Construct the vector-valued function $
			\bm{\phi}(\mathbf{s})=\left(\tau f_{\textnormal{min}}\left(\mathbf{s}\right)+\left(1-\tau\right) \left( \trans{\bm{1}}\mathbf{v}_{\nt}-f_{\textnormal{min}}\left(\widetilde{\mathbb{H}}\bm{1}-\mathbf{s}\right)\right) \right)\mathbf{v}_{\nt}+\mathbb{V}_{ 1}\mathbb{B}_{11}^{-1}\mathbf{s}$ on $\mathcal{R}\left(\widetilde{\mathbb{H}}\right)$, and the desired channel input is given by $\mathbf{X}=\bm{\phi}(\mathbf{S})$.
	\end{itemize}
	According to Appendix~\ref{app:nt-1}, the so-constructed channel input $\mathbf{X}=\bm{\phi}(\mathbf{S})$ satisfies $\widetilde{\mathbb{H}} \mathbf{X}=\mathbf{S}$, $\mathbf{X}\in [0,1]^{\nt}$, and 
	\begin{flalign}
		&	\mathbb{E}\left[\mathbf{X}\right] \nonumber\\
		=&\left( \tau \mathbb{E} \left[ f_{\textnormal{min}}(\mathbf{S})\right] + \left(1-\tau\right)\left( \trans{\bm{1}}\mathbf{v}_{\nt}-\mathbb{E} \left[ f_{\textnormal{min}}\left(\widetilde{\mathbb{H}}\bm{1}-\mathbf{S}\right)\right] \right) \right) \nonumber\\
		&\cdot \mathbf{v}_{\nt}  +\mathbb{V}_{ 1}\mathbb{B}_{11}^{-1} \mathbb{E}\left[\mathbf{S}\right] \nonumber \\
		=&\left( \mathbf{v}_{\nt}\trans{\mathbf{v}_{\nt}} +\mathbb{V}_{ 1}\trans{\mathbb{V}_{ 1}} \right)\bm{\alpha}\\
		=&\bm{\alpha},
	\end{flalign}
	i.e., feasible to the MIMO EC-OIC of rank $\nt-1$. 	\hfill\remarksymbol
\end{remark}

\section{Generalization to Bounded-Cost Constrained Channels}\label{sec:bc-oic}
In this section, we generalize the results developed in Secs. \ref{sec:3} and \ref{sec:equiv} to the MIMO OIC under another type of per-antenna intensity constraints as follows
\begin{subequations}\label{eqn:bcc}
	\begin{align}
		&0\le X_k \le 1 ,\label{eqn:bcc1}\\
		&\mathbb{E}\left[X_k\right]\le \alpha_k,~ k\in [\nt],\label{eqn:bcc2}
	\end{align}
\end{subequations}
i.e., the bounded-cost constraint in \cite{chen2022MISO}, where the constant $\alpha_k\in\left[0,1\right]$ is the ratio of the maximum allowed average intensity to the maximum allowed peak intensity. Along the same line, we refer to the MIMO OIC \eqref{eqn:model1} under the per-antenna constraints \eqref{eqn:bcc} as \textit{the MIMO BC-OIC}, whose capacity is denoted by 	$\mathsf{C}_{\textnormal{B}}\left(\widetilde{\mathbb{H}},\bm{\alpha},\sigma\right)$.
\vspace{-1em}
\subsection{Intensity Allocation at Low SNRs}\label{sec:low_snr_bcoic}
In this subsection, we are concerned with the low-SNR capacity asymptotics of the MIMO BC-OIC. Based on the existing result on the MIMO EC-OIC (see Theorem \ref{lowsnr:ec} and Corollary \ref{ecmc:max-var-bnd}), we only need to optimizing  the intensity allocation so as to maximize $	\mathrm{tr}\left(\mathbb{K}_{\mathbf{S}\mathbf{S}}\right)$ under the bounded intensity constraint \eqref{eqn:bcc2}.

\subsubsection{Optimal Intensity Allocation}
For simplicity of notation, we let $f_0(\mathbf{x})=-\const{V}_{\textnormal{max}}^{\textnormal{E}}\left(\mathbb{H},\mathbf{x}\right)$ and assume that $\alpha_1\ge \cdots \ge \alpha_{\nt}$ without loss of generality. Then the optimal intensity allocation for the MIMO BC-OIC is attained by solving the following nonsmooth convex optimization problem
\begin{equation}\label{prob:low}
	\begin{aligned}
		&\textnormal{minimize} &&f_0(\mathbf{x})\\
		&\textnormal{subject to} && \bm{0} \preccurlyeq \mathbf{x} \preccurlyeq \bm{\alpha}.
	\end{aligned}
\end{equation}

Although the objective function $f_0(\mathbf{x})$ involves the maximum function and hence is not always differentiable in the feasible region, the problem~\eqref{prob:low} can be efficiently solved by the off-the-shelf solver such as CVX.
To further reduce the computational burden in solving \eqref{prob:low}, we {\color{black}prove} a property of the optimal solution to \eqref{prob:low} in the following proposition.

\begin{proposition}\label{prop:maxvar}
	The optimal solution $\bm{x}^{\star}$ to~(\ref{prob:low}) is $\frac{1}{2}\bm{1}$ if $\alpha_{\nt} \ge \frac{1}{2}$, while otherwise $\bm{x}^{\star}$ satisfies $\bm{x}^{\star}\in [\alpha_{\nt},\min\left\{ \alpha_1,\frac{1}{2} \right\}]^{\nt}$.
	\begin{IEEEproof}
		See Appendix~\ref{app:maxvar}.
	\end{IEEEproof}	
\end{proposition}

Proposition \ref{prop:maxvar} reveals that the maximum element of the {\color{black}optimal} solution $\bm{x}^{\star}$ should be no larger than $\frac{1}{2}$, while the minimum one should be no less than $\alpha_{\nt}$. {\color{black}Because} the average intensity of the $\nt$-th transmitter is no larger than $\alpha_{\nt}$, we {\color{black}immediately see} that, to maximize $\mathrm{tr}\left(\mathbb{K}_{\mathbf{S}\mathbf{S}}\right)$, the average intensity of the $\nt$-th transmitter, i.e., the one with a minimum maximum-allowed average intensity, is exactly $\alpha_{\nt}$. The above facts help us narrow down the feasible region of the problem \eqref{prob:low} and reduce the number of variables to be optimized.

\subsubsection{Ladder Allocation}
It is noticed that a tractable form of the optimal solution $\bm{x}^{\star}$ has not been found for the general MIMO BC-OIC except for the MISO BC-OIC (or equivalently, the rank-one MIMO BC-OIC). In more details, for the MISO BC-OIC, a ladder-like intensity allocation $\mathbf{x}= \min\{\beta \bm{1},\bm{\alpha}\}$, for some $\beta\in \left[\alpha_{\nt},\min\{\alpha_1,\frac{1}{2}\} 	
\right]$, has proved to be capacity-achieving in \cite{chen2022MISO}.  Although such a ladder-like intensity allocation may no longer be optimal for general BC-OICs even at low SNRs, it exhibits robust and near-optimal performance, which will be numerically verified in Sec.~\ref{sec:numerical}. More importantly, the optimal setting of $\beta$ in the ladder-like intensity allocation can be obtained very efficiently even for the MIMO channels. 

Assume that $\alpha_1 \ge \alpha_2 \ge \cdots \ge \alpha_{\nt}$, and let $\const{V}(\beta)=\const{V}_{\textnormal{max}}^{\textnormal{E}}\left(\mathbb{H},\min\{\beta \bm{1},\bm{\alpha}\}\right)$ and $k_\beta=\max \left\{k \in [\nt]:\alpha_k\ge \beta\right\}$. When $\beta\in \left(\alpha_{{k_\beta}+1},\alpha_{k_\beta} \right]$, the objective $\const{V}(\beta)$ can be expressed as
\begin{flalign}
	\const{V}(\beta)=&\omega_{k_{\beta}}\beta \left(1-\beta\right)
	+2\left(1-\beta \right)\left(\sum_{i=1}^{k_{\beta}}\sum_{j=k_{\beta}+1}^{\nt}g_{ij}\alpha_j \right) \nonumber\\
	&	+ \underbrace{ \sum_{i=k_\beta+1}^{\nt}\sum_{j=k_\beta+1}^{\nt} g_{ij} \left(\min\{\alpha_i,\alpha_j\} -\alpha_i \alpha_j\right)}_{\const{K}}, 
\end{flalign}
where $\omega_{k_{\beta}}=\sum_{i=1}^{k_{\beta}}\sum_{j=1}^{k_{\beta}}g_{ij}$ and the quantity $\const{K}$ is a constant in the interval $\beta \in\left(\alpha_{{k_\beta}+1},\alpha_{k_\beta} \right]$. Since it is sufficient to consider points with zero derivative and end points in each interval, to find the solution $\beta^{\ast}$ that maximizes the piecewise quadratic function $\const{V}(\beta)$ only needs $O(\nt-1)$ steps. For simplicity, we refer to this suboptimal intensity allocation $\min\{\beta^{\ast} \bm{1},\bm{\alpha}\}$ as \textit{ladder allocation} scheme.

\subsection{Achievable Rate for Linear Precoding}
With a slight modification according to the bounded cost constraints \eqref{eqn:bcc2}, the linear precoding scheme proposed in Sec. \ref{subsec:linear_precoding_ecoic} can be used for the MIMO BC-OIC as well. For the BC-OIC, the intensity constraint \eqref{eq:equality_linear} is replaced by
\begin{flalign}\label{eq:inequality_linear}
	\trans{\mathbf{b}_j}\bm{\beta}_{\mathcal{I}}+c_j\le\beta_{j+r},~j \in [\nt-r],
\end{flalign}
and accordingly, the constraint \eqref{eq:17} is replaced by
\begin{subequations}
	\begin{align}
		\left\| \mathbf{b}_j \right\|_1    &\le 1 \\
		\trans{\mathbf{b}_j}\bm{\beta}_{\mathcal{I}}-\beta_{j+r}   & \le  \frac{1}{2}-\frac{1}{2}\left\| \mathbf{b}_j \right\|_1.
	\end{align}	
\end{subequations}
Along with the same approach in the proof of Proposition \ref{prop:linear_ec}, we characterize the optimized precoding vector $\mathbf{b}_j^{\star}$ in each iteration for the MIMO BC-OIC in the following proposition. 
\begin{proposition}
	For $\mathsf{t}\in \left\{1,2\right\}$, denote the solution and the optimal value of the following linear programming problem
	\begin{subequations}\label{prob:iter3}
		\begin{align}
			\textnormal{max} ~~&\quad\quad 
			\trans{\mathbf{w}}_{\mathsf{t}}\bm{\mu}_{\mathsf{t}}
			\\
			\textnormal{s.t.} ~~
			& \quad\quad \bm{\mu}_{\mathsf{t}}\succcurlyeq \bm{0} \label{eq:24b}\\
			&\quad\quad \trans{\bm{1}}\bm{\mu}_{\mathsf{t}}\le 1, \label{eq:24c} \\
			&\quad\quad \trans{\bm{\alpha}_{\mathsf{t}}}\bm{\mu}_{\mathsf{t}}\le \alpha_{j+r},  \label{eq:24d}
		\end{align}
	\end{subequations}
	by $\bm{\mu}_{\mathsf{t}}^{\star}$ and $\const{L}_{\mathsf{t}}^{\star}$. If $1+\trans{\mathbf{w}}_{1}\bm{\mu}_{1}\ge \left|1-\trans{\mathbf{w}}_{2}\bm{\mu}_{2}\right|$, then $\mathbf{b}_{j,\mathcal{K}}^{\star}=\bm{\mu}_1$ and $\mathbf{b}_{j,\mathcal{K}^{\text{c}}}^{\star}=\bm{0}$; otherwise, $\mathbf{b}_{j,\mathcal{K}^{\text{c}}}^{\star}=\bm{\mu}_2$ and $\mathbf{b}_{j,\mathcal{K}}^{\star}=\bm{0}$.
\end{proposition}
\begin{IEEEproof}
	Omitted.
\end{IEEEproof}

Denote $\mathbb{B}_{\mathsf{B}}^{\dag}$ as the suboptimal precoding matrix by using the aforementioned iterative method with corresponding modification for the MIMO BC-OIC. Let $\const{D}_{\mathsf{B}}=\left| \det \left( \mathbb{H}_{\mathcal{I}}+\mathbb{H}_{\mathcal{I}^{c}}\mathbb{B}^{\dag}_{\mathsf{B}}  \right)\right| $. Then an achievable rate for the linear precoding scheme  given by the following proposition.
\begin{proposition}\label{prop:linear_precoding2}
	The capacity of the MIMO optical intensity channel \eqref{eq:linear_transmission} with linear precoding and under bounded-cost constraints \eqref{eqn:bcc} is lower-bounded as 
	\begin{flalign}\label{eq:linear_precoding2}
		&\mathsf{C}_{\mathsf{B}}\left(\widetilde{\mathbb{H}},\bm{\alpha},\sigma\right) \nonumber \\
		\ge&
		\frac{ r }{2}\log\left(1+
		\frac{\exp\left( 2\const{D}_{\mathsf{B}}\left(\sum_{i=1}^{r}\mathfrak{h}_{\text{TE}}\left(\min\left\{\alpha_i,\frac{1}{2}\right\}\right)\right)/r \right)}{2\pi e \sigma^2}\right).
	\end{flalign}
\end{proposition}
\begin{IEEEproof}
	The proposition is concluded by combining the proof technique for Proposition \ref{prop:linear_precoding} and the fact that the differential entropy $\mathfrak{h}_{\text{TE}}(\alpha)$ is maximized when $\alpha=\frac{1}{2}$.
\end{IEEEproof}

%

\subsection{Achievable Rate by Optimal Rank-One Transmission}\label{sec:rank-one}
In Sec. \ref{sec:qr-sic}, a QR-SIC receiver is presented for the MIMO EC-OIC. The key idea is flipping the inputs of the last $\nt-r+ 1$ transmitters that corresponding to negative channel gains $p_{r,k}$ (in the sense of the transformed channel matrix $\mathbb{P}$ in \eqref{eq:31}) for $k\in \left\{ r,r+1,\cdots,\nt \right\}$, and hence, the flipped subchannel is exactly a MISO EC-OIC. Unfortunately, such a method can not be generalized to the bounded-cost case.
Nevertheless, Thm. \ref{thm:rankone} shows that the first eigen-subchannel $\widetilde{Y}_i=\sigma_1 \trans{\mathbf{v}}_1 \mathbf{X} +\widetilde{Z}_1$ with respect to the transformed channel \eqref{eq:equiv} can be always viewed as a MISO OIC, regardless of the form of input constraints. Implicitly, we conclude the following capacity lower bound.
\begin{proposition}\label{prop:bc_oic_rankone}
	The achievable rate of the MIMO optical intensity channel \eqref{eq:linear_transmission} either under either equal-cost constraints \eqref{eqn:ecc} or bounded-cost constraints \eqref{eqn:bcc} is lower-bounded as 
	\begin{flalign}\label{eq:rank_one_lbd}
		\mathsf{C}_{\mathsf{E} \text{ or }\mathsf{B}}\left(\widetilde{\mathbb{H}},\bm{\alpha},\sigma\right) 
		\ge
		\mathsf{C}_{\mathsf{E} \text{ or }\mathsf{B}}\left(\sigma_1\trans{\mathbf{v}}_1, \bm{\alpha},\sigma\right).
	\end{flalign}
\end{proposition}
\begin{IEEEproof}
	The bound is simply achieved by ignoring other subchannels corresponding smaller eigenvalues.
\end{IEEEproof}

The above theorem characterizes the achievable rate of only using the strongest eigen-subchannel. From the Eckart-Young theorem, we immediately know that the channel vector of the strongest eigen-subchannel is the closest rank-one matrix to the original channel matrix in the Frobenius distance. For this reason, we refer to the transmission scheme using only the strongest eigen-subchannel as \textit{optimal rank-one transmission}.
The lower bound \eqref{eq:rank_one_lbd} may be more suitable for a line-of-sight (LOS) optical communication link with an ill-conditioned channel matrix and at low-to-moderate SNRs. At high SNRs, the lower bound \eqref{eq:rank_one_lbd} will be loose since its pre-log factor is limited to ${1}/{2}$ rather than ${r}/{2}$ (e.g., full multiplexing gains).

\vspace{-1em}
\subsection{Achievable Rate Based on General EPI}
Same as in Sec. \ref{sec:gepi_ec}, we present a general-EPI-based capacity lower bound for the MIMO BC-OIC in the following proposition.

\begin{proposition}\label{prop:bc_oic_gepi}
	The capacity of the MIMO optical intensity channel \eqref{eq:linear_transmission} and under bounded-cost constraints \eqref{eqn:bcc} is lower-bounded as 
	\begin{flalign}
	&\mathsf{C}_{\textsf{B}}\left(\widetilde{\mathbb{H}},\bm{\alpha},\sigma\right) \nonumber \\
	\ge&
	\frac{r}{2}\log\left(1+\frac{\sqrt[r]{\det\left(\sum_{i=1}^{\nt}\exp(2\mathfrak{h}_{\text{TE}}(\min\left\{ \alpha_i,\frac{1}{2}\right\})) \cdot\left( \mathbf{h}_i\trans{\mathbf{h}}_i\right)\right)} }{2\pi e \sigma^2}\right).
	\end{flalign}
	\begin{IEEEproof}
	The proposition is concluded by combining Proposition \ref{prop:gepi_ec} and the fact that the differential entropy $
	\mathfrak{h}_{\text{TE}}(\alpha)$ of the truncated exponential distribution is maximized at $\alpha=\frac{1}{2}$.
	\end{IEEEproof}
\end{proposition}

\subsection{Capacity Lower Bound for Rank-$(\nt-1)$ Case Based on Maximum Entropy}

Although no similar capacity expression has been found for the MIMO BC-OIC of rank $\nt-1$, we can investigate its capacity by maximizing $\mathsf{C}_{\textnormal{E}}\left(\widetilde{\mathbb{H}},\bm{x},\sigma\right)$ over all average-intensity vectors $\mathbf{x}$ satisfying $\bm{0} \preccurlyeq \mathbf{x} \preccurlyeq \bm{\alpha}$. Thus, by maximizing the maximum differential entropies $\gamma_{ \textnormal{E}}$ in \eqref{prob:max_ent} over all feasible average-intensity vectors and based on Sion's minimax theorem, the following theorem presents an achievable rate of the MIMO BC-OIC of rank $\nt-1$ that is asymptotically capacity-achieving at high SNRs.
\begin{theorem}
	\label{thm:highsnr_BCMC}
	The capacity of the MIMO BC-OIC of rank $r=\nt-1$ is lower-bounded by
\begin{equation}
	\mathsf{C}_{\mathsf{B}}\left(\widetilde{\mathbb{H}},\bm{\alpha},\sigma\right)
	\ge
	\frac{ \nt-1 }{2}\log\left(1+
	\frac{\exp\left( 2\gamma_{\mathsf{B}}/\left(\nt-1 \right)\right)}{2\pi e \sigma^2}\right),
\end{equation}
	where 
	\begin{flalign} \label{eq:gammaB}
		\gamma_{ \mathsf{B}} = 	& \min_{\nu \in \mathbb{R}, \mathbf{u}\in\mathbb{R}^r \atop \lambda_1,\lambda_{2}\ge 0}
		\left(\sum_{i=1}^{\nt}\alpha_i\cdot \max\left\{  \trans{\mathbf{u}}\widetilde{\mathbf{h}}_i+\left(\lambda_1-\lambda_2 \right)v_{\nt,i},0\right\} \right)
		\nonumber\\
		&+  \left(\trans{\bm{1}}{\mathbf{v}}_{\nt}\right)\lambda_2-1-\nu
		\nonumber\\
		+& \int_{\mathcal{R}\left(\widetilde{\mathbb{H}}\right)}  \exp \left(\nu -\trans{\mathbf{u}}\mathbf{s}-
		\lambda_1 f_{\textnormal{min}}\left(\mathbf{s}\right)-\lambda_2 f_{\textnormal{min}}\left(\widetilde{\mathbb{H}}\bm{1}-\mathbf{s}\right)
		\right) \dd \mathbf{s}.
	\end{flalign}
		is the maximum differential entropy of the equivalent input $\mathbf{S}$ for the MIMO BC-OIC. 	
\end{theorem}
\begin{IEEEproof}
	The maximum differential entropy of the input for the MIMO BC-OIC of rank $\nt-1$ is given by the following max-min problem:
	\begin{flalign} 
		\gamma_{ \textnormal{B}} = 	&\max_{\bm{0} \preccurlyeq \mathbf{x} \preccurlyeq \bm{\alpha}} \min_{\nu \in \mathbb{R}, \mathbf{u}\in\mathbb{R}^r \atop \lambda_1,\lambda_{2}\ge 0}
		\trans{\mathbf{u}}\widetilde{\mathbb{H}}\bm{x}+\lambda_1 \trans{\mathbf{v}}_{\nt}\bm{x}
		+\lambda_2  \trans{\mathbf{v}}_{\nt} \left(\bm{1}-\bm{x}\right)-1-\nu
		\nonumber\\
		+& \int_{\mathcal{R}\left(\widetilde{\mathbb{H}}\right)}  \exp \left(\nu -\trans{\mathbf{u}}\mathbf{s}-
		\lambda_1 f_{\textnormal{min}}\left(\mathbf{s}\right)-\lambda_2 f_{\textnormal{min}}\left(\widetilde{\mathbb{H}}\bm{1}-\mathbf{s}\right)
		\right) \dd \mathbf{s}.
	\end{flalign}
	Clearly, when fixing $\mathbf{x}$ the above objective function is convex with respect to variables $\nu$, $\mathbf{u}$, $\lambda_1$, and $\lambda_{2}$, while the objective function is affine with respect to variables $\mathbf{x}$ when fixing $\nu$, $\mathbf{u}$, $\lambda_1$, and $\lambda_{2}$. Note that the feasible region of the outer maximization problem is a box, and hence, convex and compact, while that of the inner minimization problem is the intersection of two closed half-spaces, and hence, convex as well. For this reason, Sion's minimax theorem enable us to exchange the maximum operation and the minimum one. The entropy
	\eqref{eq:gammaB} can be easily obtained by noting that, under a separable box constraint $0 \le x_i \le \alpha_i$ for $i\in [\nt]$, the inner product $\left(\trans{\mathbf{u}}\widetilde{\mathbb{H}}+\lambda_1 \trans{\mathbf{v}}_{\nt}-\lambda_2  \trans{\mathbf{v}}_{\nt} \right)\bm{x}$ is maximized by letting $x_i=\alpha_i$ when $\trans{\mathbf{u}}\widetilde{\mathbf{h}}_i+\left(\lambda_1-\lambda_2 \right)v_{\nt,i}>0$ and $x_i=0$ otherwise. 
\end{IEEEproof}

\section{Numerical Results}\label{sec:numerical}
In this section, we present numerical evaluation of our derived capacity results. For fairness purposes, both the indoor VLC channel and the outdoor lognormal fading channel are considered. Due to the normalized maximum allowed peak intensity, we define the signal-to-noise ratio as $\text{SNR}\triangleq {1}/{\sigma}$.

\subsection{Indoor VLC Scenario}
\begin{figure}[hbtp!]
	\centering
	\resizebox{7cm}{!}{\includegraphics{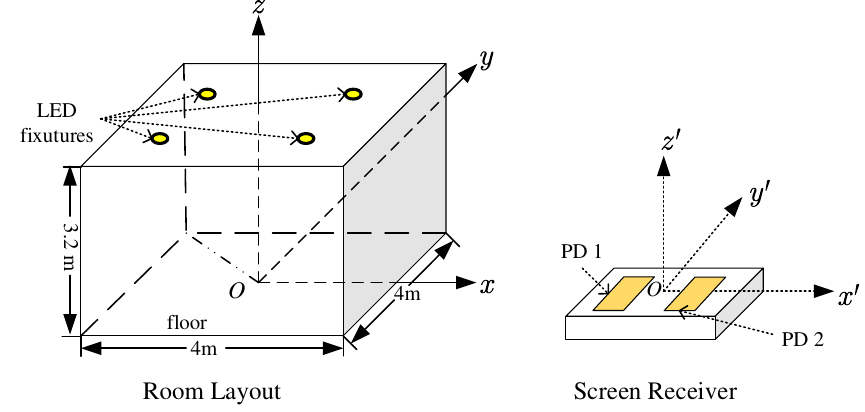}}
	\centering \caption{Room layout and receiver structure. Four LED fixtures are fixed in the ceiling of a $4\, \text{m} \times 4\, \text{m} \times 3.2 \,\text{m} $ room with LED coordinates (in the $xyz$ system) $\left(\pm 1, \pm 1, 3.2\right)$  (unit: m), and two PDs are placed on one side of the user equipment with PD coordinates (in the $x'y'z'$ system) $(\pm 0.05, 0 ,0 )$ (unit: m). }
	\label{fig:room}
\end{figure}

We consider a four-input two-output ($4 \times 2$) VLC system in a $4\, \text{m} \times 4\, \text{m} \times 3.2 \,\text{m} $ room. The layout of the room and the structure of PD receivers are illustrated in Fig. \ref{fig:room}. Lambertian order of four LEDs and the field of view of the receiver PDs are set to $1$ and $60$ deg\footnote{We assume two PDs are of a same specification, and due to scaling we do not assign PD parameters that are linear with channel gains (e.g., area, responsivity, refractive index, etc.).}. For simplicity, we only take LOS channel gains into account, which are computed by Lambert's model \cite{Soltani2019TCOM}. Without loss of generality, the channel matrix is scaled by a constant factor such that the maximum entry of all possible channel matrices is $1$. 

\vspace{-1em}
Due to the requirement of dimming control and color adjustment in the indoor VLC scenario, the equality constraints \eqref{eqn:ecc} are imposed on the LED transmitters. 
In Fig.~2, we plot our derived capacity bounds, existing upper bounds \eqref{eq:max-var-bnd} and \cite[Thm. 17]{limoserwangwigger20_1}, and the mutual information achieved by the maximal correlated multivariate distribution proposed in \cite{chaabanrezkialouini18_2} for this $4\times 2$ EC-OIC with $\bm{\alpha}=\trans{(0.4,0.4,0.3,0.2)}$ when the receiver locates at four different points $O: (0,0,1)$, $P_1: (1,0,1)$, $P_2: (1,1,1)$, and $P_3: (2,2,1)$ (unit: m) in the $xyz$ system. It suggests that: 1) The maximally correlated multivariate distribution achieves the channel capacity at low SNRs; 2) Simply using the optimal rank-one transmission 
outperforms other lower bounds at moderate SNRs since channel matrices of the indoor VLC are usually highly ill-conditioned; and 3) Achievable rates of the QR-SIC scheme and the linear precoding scheme are both close to the upper bound within 1 nats per channel use.

\begin{figure*}[!htbp]
	\centering
	\label{fig:indoor} 
	\begin{minipage}[!htbp]{0.5\textwidth} 
		\centering 
		\includegraphics[width=8cm]{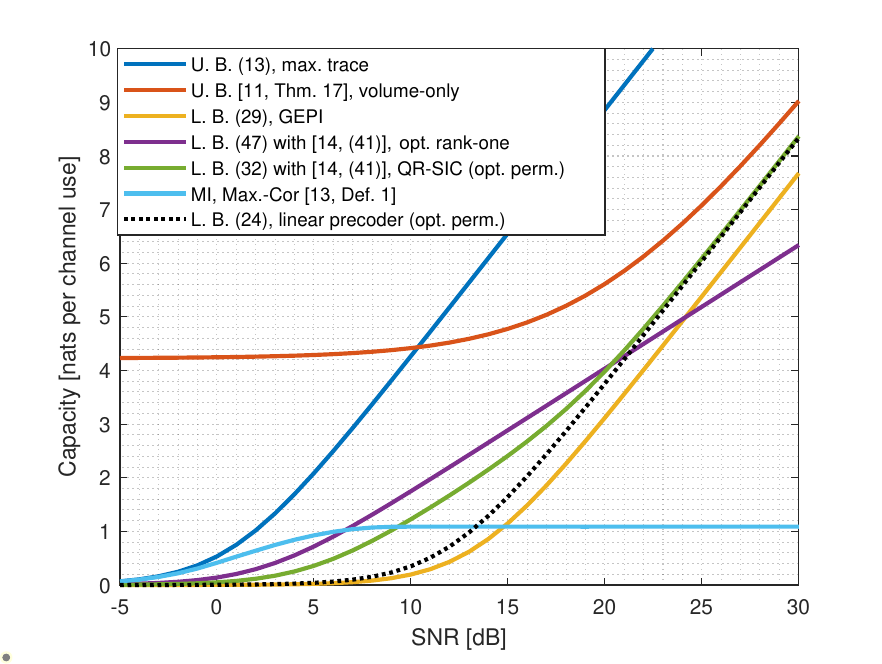} 
	\end{minipage}%
	\begin{minipage}[!htbp]{0.5\textwidth} 
		\centering 
		\includegraphics[width=8cm]{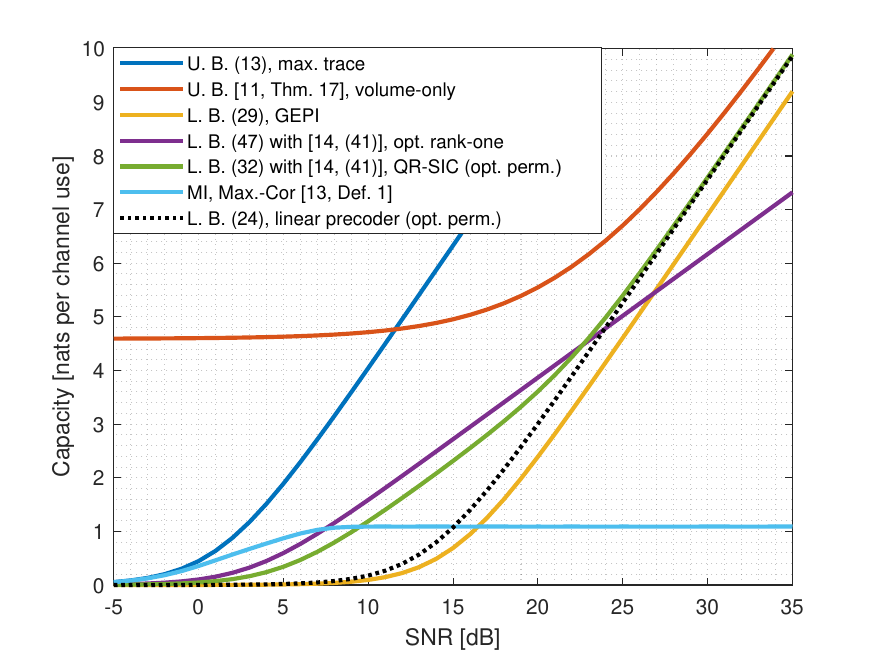} 
	\end{minipage}%

	\begin{minipage}[!htbp]{0.5\textwidth} 
		\centering 
		\includegraphics[width=8cm]{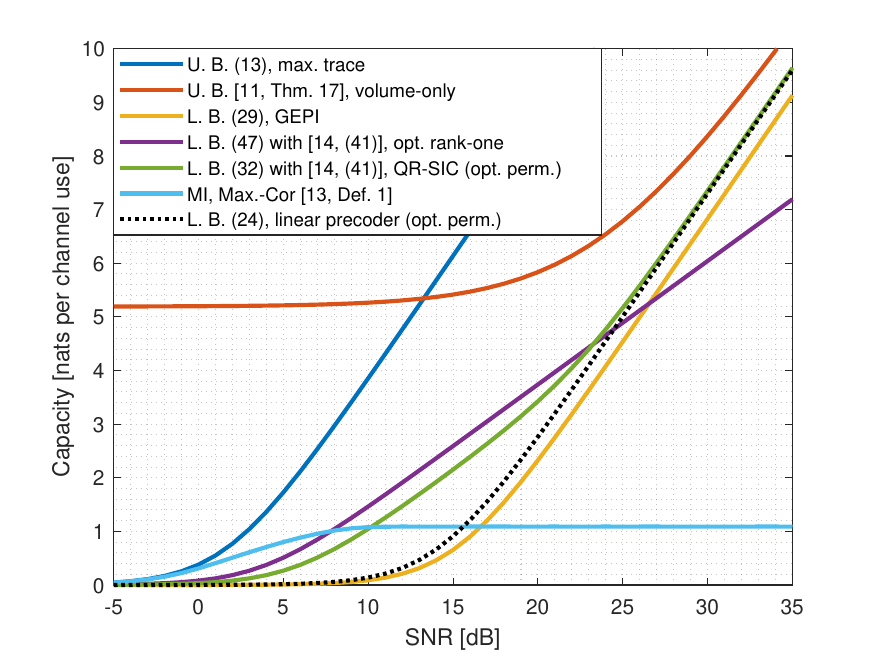} 
	\end{minipage}%
	\begin{minipage}[!htbp]{0.5\textwidth} 
		\centering 
		\includegraphics[width=8cm]{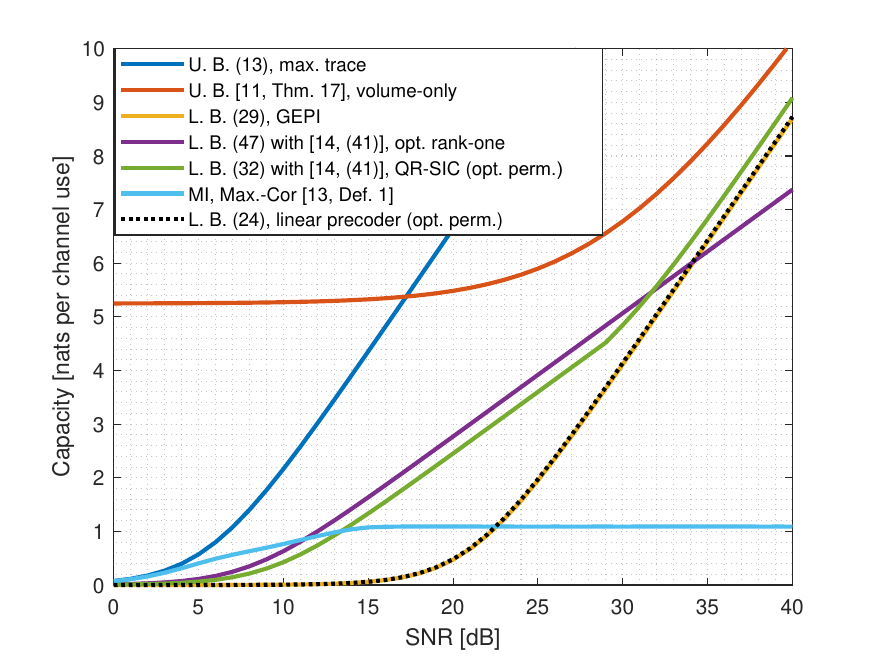} 
	\end{minipage}%
	\centering \caption{Capacity bound with the receiver locating at $O$ (upper left), $P_1$ (upper right), $P_2$ (lower left), and $P_3$ (lower right). The involved capacity lower bounds \eqref{eq:linear_precoding} and \eqref{eq:ec_oic_qr} are optimized over all possible permutations of transmitters.} 
	\vspace{-0.3cm}
\end{figure*}

\vspace{-1em}

\subsection{Outdoor Fading Channel}
Next, we turn to the outdoor free-space optical communication with limited optical power, which is modeled by the MIMO BC-OIC. For simplicity, the entries $h_{ij}$ of fading channel matrices are assumed to be independent and identically lognormal distributed, i.e., $h_{ij}=\exp(N_{ij})$ with i.i.d. $N_{ij}\sim \mathcal{N}(0,1)$.

In Fig.~3, for $6\times 4$ lognormal fading channels with the bounded average intensities $\bm{\alpha}=\trans{(0.4,0.4,0.3,0.3,0.2,0.2)}$, we plot the bounds on the ergodic capacity by averaging capacity bounds over channel realizations. It is shown that: 1) Our proposed suboptimal intensity allocation (i.e., the ladder allocation in Sec. \ref{sec:low_snr_bcoic}) approaches the ergodic capacity at low SNRs; 2) The averaged achievable rate of our proposed linear precoding scheme is higher than the GEPI-based lower bound at all SNRs and close to the loose upper bound within 1 nats per channel use; and 3) The performance gain of the optimal rank-one transmission is limited at moderate SNRs since ignoring other eigen-subchannels of the considered channel matrices with i.i.d. fading coefficients leads to considerable loss.

\begin{figure}[hbtp!]
	\centering
	\label{fig:BOIC_capacity}
	\includegraphics[width=8cm]{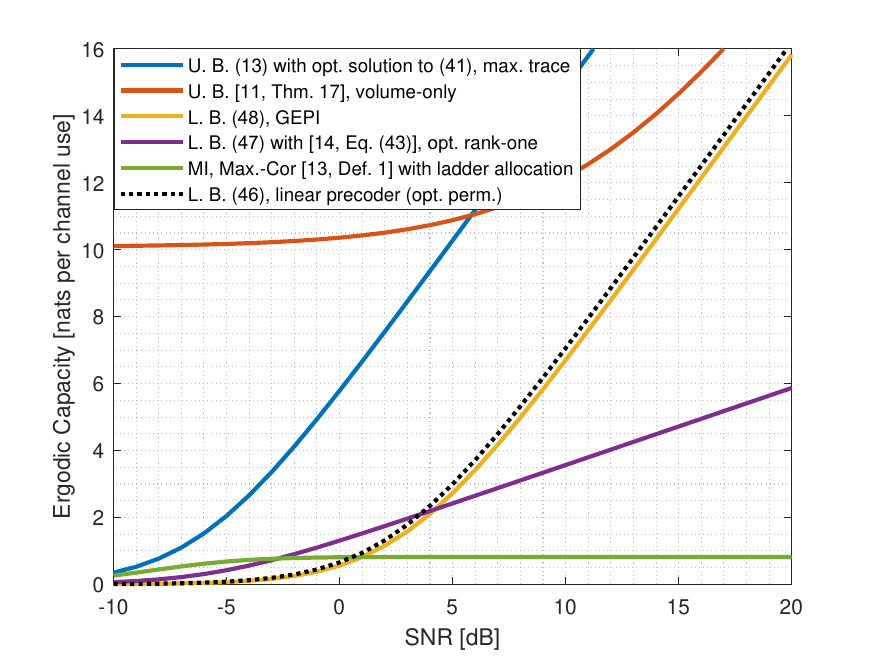}
	\centering \caption{Bounds on the ergodic capacity of a $6\times 4$ MIMO BC-OIC with $\bm{\alpha}=\trans{(0.4,0.4,0.3,0.3,0.2,0.2)}$ and lognormal fading. The involved lower bounds \eqref{eq:linear_precoding2} are optimized over all possible permutations of transmitters.} 
\end{figure}
\vspace{-1em}


%

\subsection{Results for OICs of Rank $n_t-1$}
\begin{figure}[!htbp]
	\vspace{-3mm}
	\centering
	
	\begin{minipage}{8cm}
		\centering\includegraphics [width=8cm]{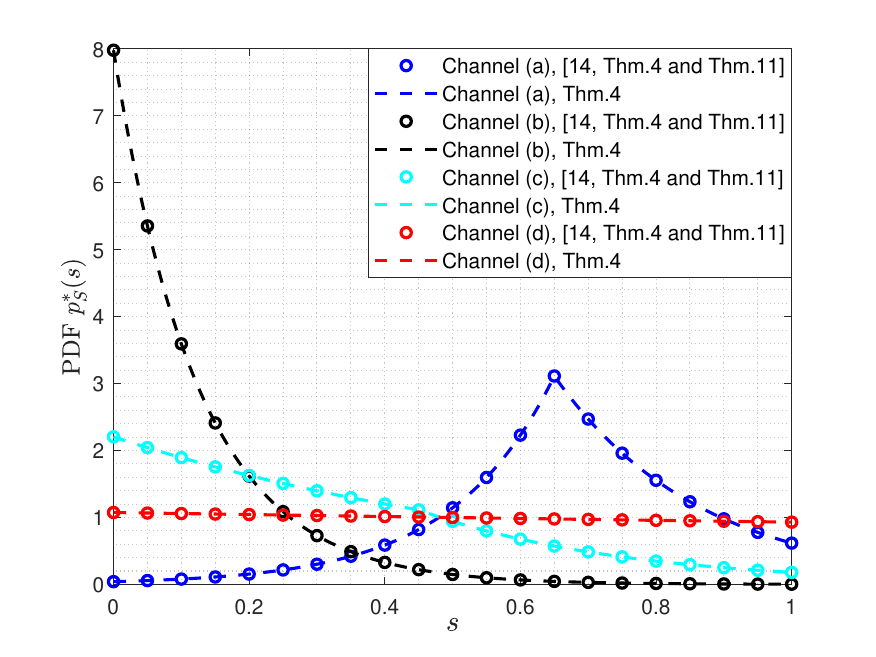}
	\end{minipage}
	\begin{minipage}{8cm}
		\centering\includegraphics [width=8cm]{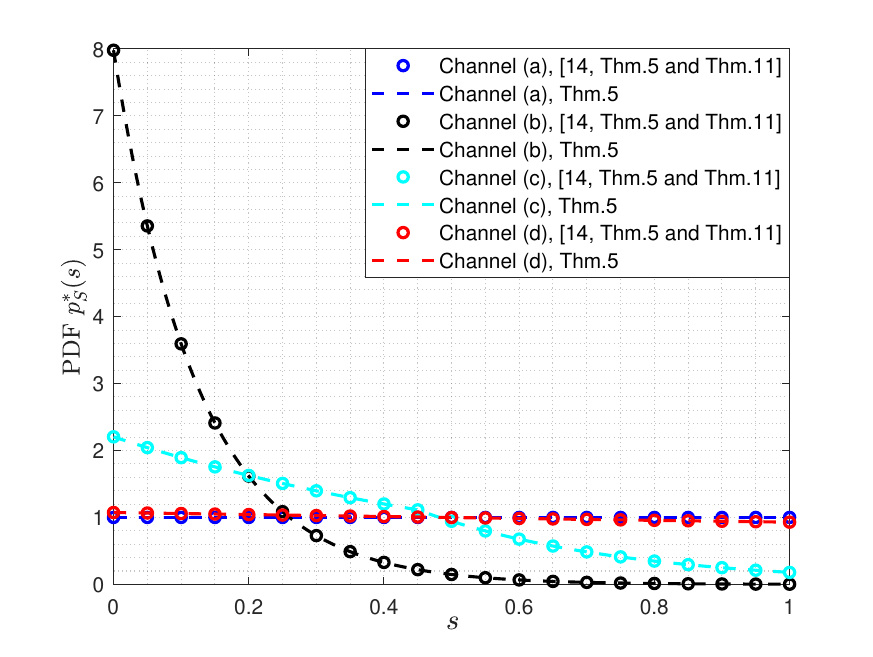}
	\end{minipage}
	
	\vspace{-3mm}
	\caption{Probability density functions of {\color{black}maximum-entropy distributions for $2\times 1$ EC-OICs (upper) and $2\times 1$ BC-OICs (lower)}.} \label{fig:pdf}
\end{figure}

To verify our derived results on the EC-OIC and the BC-OIC of rank $\nt-1$, we specially look at $2\times1$ EC- and BC-OICs, which are known to be of either rank-one or rank-($\nt-1$). For four $2\times 1$ OICs (parameterized in Table \ref{tab:channel}), Fig. \ref{fig:pdf} plots probability density functions of maximum-entropy distributions,
whose parameters are numerically computed by using results in \cite{chen2022MISO} (i.e., treating considered channels as the MISO channels) and Thms. \ref{thm:highsnr_ECMC} and \ref{thm:highsnr_BCMC} in this paper (i.e., treating those as MIMO channels of rank $\nt-1$), respectively. It can be seen that the numerical results obtained by both methods show a very good match, which reflects the correctness of our theoretic results for the special case of $r=\nt-1$.

\begin{table}[!htbp]
	\caption{Channel Parameters.}
	\centering\footnotesize{
		\begin{tabular}{|c|c|c|c|c|}
			\hline
			\xrowht{8pt}
			Channel&$\mathbb{H}$&$\trans{\bm{\alpha}}$&$\gamma_{ \textnormal{E}}$&$\gamma_{ \textnormal{B}}$\\ 
			\hline
			(a)&$\left(0.65,0.35 \right)$&$\left(0.90,0.20 \right)$&$-0.3780$&$0.0000$\\
			\hline
			(b)&$\left(0.25,0.75 \right)$&$\left(0.20,0.10 \right)$&$-1.0798$&$-1.0798$\\
			\hline
			(c)&$\left(0.45,0.55 \right)$&$\left(0.60,0.10 \right)$&$-0.1978$&$-0.1979$\\
			\hline
			(d)&$\left(0.20,0.80 \right)$&$\left(0.52,0.48 \right)$&$-0.0009$&$-0.0009$\\
			\hline
	\end{tabular}}\label{tab:channel}
\end{table}

\vspace{-2em}


\section{Conclusion and Discussion}\label{sec:conclusion}
In this paper, we investigate the capacities of MIMO OICs under two different types of per-antenna intensity constraints. Based on a channel reduction method, we shows the positivity of channel gains of the strongest eigen-subchannel for the general MIMO OIC, which motivates the optimal rank-one transmission scheme. Then we derive various capacity bounds by analyzing achievable rates of different transceiver frameworks and some standard information-theoretic tools. Besides, an equivalent expression is given in a special case that the rank of the channel is one less than the number of transmit antennas. The developed results are  generalized to the MIMO OIC with bounded-cost constraints.

So far, for the MIMO OIC under per-antenna intensity constraints, equivalent capacity expressions converting the original linear channel into the vector Gaussian channel with moment constraints are characterized only for two special cases, i.e., the rank-one case in \cite{chen2022MISO} and the rank-$(\nt-1)$ case in this paper. Characterization of its capacity in general case is still an open problem. The main difficulty lies in the description of the moment constraints on the transformed vector Gaussian channel. To develop a unified theoretical framework for general MIMO EC- and BC-OIC of arbitrary rank may be an interesting research direction in the future.

%
%
%
%


\vspace{-0.5em}
\appendices

\section{Proof of Proposition \ref{prop:linear_ec}}\label{app:linear_ec}
	The $\ell_1$-norm maximization problem \eqref{prob:iter} can be solved by respectively maximizing $\trans{\mathbf{b}}_j\mathbf{w}$ and $-\trans{\mathbf{b}}_j\mathbf{w}$ under the constraints \eqref{eq:20b} and \eqref{eq:20c}. We first consider the problem of maximizing $\trans{\mathbf{b}}_j\mathbf{w}$. It can be easily verified that the feasible set always contains $\bm{0}$, and therefore is not empty. For any feasible point $\check{\mathbf{b}}=\trans{\left(\check{b}_1,\cdots,\check{b}_r \right)}$, the inequality $\trans{\mathbf{w}}\check{\mathbf{b}}\le \trans{\mathbf{w}}\check{\mathbf{b}}^{\dag}$ holds, where $\check{b}^{\dag}_{k}=0$ if $k\in \mathcal{K}^{\text{c}}$ and otherwise $\check{b}^{\dag}_{k}=\max\left\{  \check{b}_{k},0\right\}$. Next we notice that $\check{\mathbf{b}}^{\dag}$ is also feasible based on two facts: 1) it is clear that $\left\| \check{\mathbf{b}}^{\dag} \right\|_1 \le \left\| \check{\mathbf{b}} \right\|_1$, and hence, the constraint \eqref{eq:20b} is satisfied; 2) The variation on the left hand side of \eqref{eq:20c} satisfies 
	\begin{flalign}
		&\left|\left( \left|	\trans{\bm{\beta}_{\mathcal{I}}}\check{\mathbf{b}}_j^{\dag}-\beta_{j+r} \right| - \left|	\trans{\bm{\beta}_{\mathcal{I}}}\check{\mathbf{b}}_j-\beta_{j+r} \right|\right)\right| \nonumber \\
		\le& \sum_{k\in \mathcal{K}}\left| \beta_{k} \check{b}_k  \right|+\sum_{k\in \mathcal{K}^{\text{c}}}\left| \beta_{k} \check{b}_k  \right|1_{\mathbb{R}_+}(-\check{b}_k),
	\end{flalign}
	while that on the right hand side satisfies
	\begin{flalign}
		&	\frac{1}{2}-\frac{1}{2}\left\| \check{\mathbf{b}}_j^{\dag} \right\|_1-\left(\frac{1}{2}-\frac{1}{2}\left\| \check{\mathbf{b}}_j \right\|_1\right)\nonumber\\
		=& \frac{1}{2}\left( \sum_{k\in \mathcal{K}}\left|  \check{b}_k  \right|+\sum_{k\in \mathcal{K}^{\text{c}}}\left| \check{b}_k  \right|1_{\mathbb{R}_+}(-\check{b}_k) \right) \nonumber \\
		\ge & \sum_{k\in \mathcal{K}}\left| \beta_{k} \check{b}_k  \right|+\sum_{k\in \mathcal{K}^{\text{c}}}\left| \beta_{k} \check{b}_k  \right|1_{\mathbb{R}_+}(-\check{b}_k),\label{eq:22}
	\end{flalign}
	where \eqref{eq:22} follows from the fact that $\left| \beta_k \right| \le \frac{1}{2}$.
	Hence, the inequality \eqref{eq:20c} also holds for $\check{\mathbf{b}}_j^{\dag}$.Thus we know the optimal solution $\mathbf{b}^{\star}$ that maximizes $\trans{\mathbf{b}}_j\mathbf{w}$ must be nonnegative, and moreover ${b}^{\star}_{k}=0$ for all $k\in \mathcal{K}^{\text{c}}$. By using this property, the inequalities \eqref{eq:20b} and \eqref{eq:20c} can be converted into \eqref{eq:21d} and \eqref{eq:21e}.
	The same technique applies to the another problem of maximizing $-\trans{\mathbf{b}}_j\mathbf{w}$ as well. The proposition is concluded by combining optimization results for maximizing $\trans{\mathbf{b}}_j\mathbf{w}$ and $-\trans{\mathbf{b}}_j\mathbf{w}$.

\section{Proof of Theorem~\ref{thm:nt-1}}\label{app:nt-1}

The following lemma reveals that any feasible input $\mathbf{X}$ can be seen as some function of its image $\mathbf{S}=\widetilde{\mathbb{H}}\mathbf{X}$ when the cost is only determined by the average intensity of the channel input. 

\begin{lemma}\label{lemma:input}
	Let $\mathcal{X}_{\textnormal{conv}}$ be a convex and closed set in  $\mathbb{R}^{\nt}$. For any input $\mathbf{X}$ with $\supp\,\mathbf{X}\subseteq \mathcal{X}_{\textnormal{conv}}$, there exists an input $\breve{\mathbf{X}}$, which is a function of $\vect{S}$,  satisfying
	\begin{enumerate}
		\item $\supp\,\breve{\mathbf{X}}\subseteq \mathcal{X}_{\textnormal{conv}}$;
		\item $\mathbb{E}\left[\mathbf{X}\right]=\mathbb{E}\left[\breve{\mathbf{X}}\right]$;
		\item $\vect{S} = \widetilde{\mathbb{H}}\mathbf{X} \overset{\textnormal{d}}{=} \widetilde{\mathbb{H}}\tilde{\mathbf{X}}$.
	\end{enumerate} 
\end{lemma}
\begin{IEEEproof}
	Construct a vector-valued function $\bm{\psi}: \supp\,\mathbf{S} \to \mathcal{X}_{\text{conv}}$ as $
	{\bm{\psi}}(\mathbf{s})=\mathbb{E}\left[\mathbf{X}|\widetilde{\mathbb{H}}\mathbf{X}=\mathbf{s}\right]$.
	It is straightforward that $\widetilde{\mathbb{H}}\cdot{\bm{\psi}}(\mathbf{s})=\mathbf{s}$. Since the intersection of an affine set and $\mathcal{X}_{\text{conv}}$ is convex and closed, we have ${\bm{\psi}}(\mathbf{s}) \in  \mathcal{X}_{\text{conv}}$. 
	The proof is concluded by letting $\breve{\mathbf{X}}={\bm{\psi}}(\mathbf{S})$, whose expectation satisfies $\mathbb{E}\left[\breve{\mathbf{X}}\right]=\mathbb{E}\left[{\mathbf{X}}\right]$ due to the law of total expectation.
\end{IEEEproof}

Since the zonotope $\mathcal{R}\left({\widetilde{\mathbb{H}}}\right)$ is convex and closed, it is sufficient to let the channel input $\mathbf{X}$ be a function of a random vector $\mathbf{S}$ with $\supp\, \mathbf{S}\subseteq \mathcal{R}\left({\widetilde{\mathbb{H}}}\right)$ and under constraints specified in the following.

For any $\mathbf{s}  =\trans{\left(s_1,\, s_2, \, \cdots, s_{\nt-1}  \right)}  \in \mathcal{R}\left({\widetilde{\mathbb{H}}}\right)$, we define 
$$\mathcal{X}_{\mathbf{s}}\triangleq \left\{ \mathbf{x}\in [0,1]^{\nt}: \widetilde{\mathbb{H}}\mathbf{x}=\mathbf{s} \right\}.$$
It is straightforward that $\mathcal{X}_{\mathbf{s}}$ must be nonempty and convex. Since the matrix $ \widetilde{\mathbb{H}}$ has full row rank, its Moore-Penrose pseudoinverse is given by
$
\widetilde{\mathbb{H}}^+=\trans{\widetilde{\mathbb{H}}}
\left(\widetilde{\mathbb{H}} \trans{\widetilde{\mathbb{H}}} \right)^{-1}
=\mathbb{V}_{ 1}\mathbb{B}_{11}^{-1}
$. Note that the set $\mathcal{X}_{\mathbf{s}}$ is the intersection of the affine set $ \text{null}(\widetilde{\mathbb{H}})+\widetilde{\mathbb{H}}^+\mathbf{s}$ and the unit $\nt$-cube, where the null space $\text{null}(\widetilde{\mathbb{H}})=\left\{\lambda \mathbf{v}_{\nt}:\lambda\in\mathbb{R} \right\}$ since $\textnormal{rank}\left(\widetilde{\mathbb{H}}\right)=\nt-1$ and $\mathbf{v}_{\nt}$ is orthogonal to $\mathbb{V}_{ 1}$. Hence, $\mathcal{X}_{\mathbf{s}}$ is a one-dimensional convex hull, i.e., a line segment, and can be expressed in the form of $
	\mathcal{X}_{\mathbf{s}}
	=\left\{ \lambda \mathbf{v}_{\nt}+\mathbb{V}_{ 1}\mathbb{B}_{11}^{-1}\mathbf{s}:
	\lambda\in \left[  f_{\textnormal{min}}(\mathbf{s}),
	f_{\textnormal{max}}(\mathbf{s})\right]
	\right\}$, where two real-valued functions  $ f_{\textnormal{min}}(\mathbf{s})$ and $ f_{\textnormal{max}}(\mathbf{s})$ will be characterized later. 

Note that, for any $i\in [\nt]$, the $i$-th element of $\lambda \mathbf{v}_{\nt}+\mathbb{V}_{ 1}\mathbb{B}_{11}^{-1}\mathbf{s}$ is
$
\kappa_{i}(\lambda)=\lambda v_{\nt,i}+\sum_{j=1}^{\nt-1}  v_{j,i} s_j/\sigma_j$,
where $v_{j,i}$ stands for the $i$-th element of $\mathbf{v}_j$. 

We first study the case of $v_{\nt,i}= 0$. Note that, for any $\mathbf{s}\in \mathcal{R}\left({\widetilde{\mathbb{H}}}\right)$, there exists an $\mathbf{x}\in [0,1]^{\nt}$ such that
\begin{flalign}
	\widetilde{\mathbb{H}}^{+}\mathbf{s}&=\mathbb{V}_{ 1}\trans{\mathbb{V}}_{  1}\mathbf{x} \nonumber  \\
	&=\left(\mathbb{I}_{\nt}-\mathbf{v}_{ \nt}\trans{\mathbf{v}}_{\nt}\right)\mathbf{x} \label{eq:60}\\
	&=\mathbf{x}-\mathbf{v}_{ \nt}\trans{\mathbf{v}}_{\nt}\mathbf{x},
\end{flalign}
where $\mathbb{I}_{\nt}$ denotes the identity matrix of size $\nt\times \nt$ and~\eqref{eq:60} follows from the equality
\begin{equation}\label{eq:62}
	\mathbb{I}_{\nt}=\mathbb{V}\trans{\mathbb{V}}=\mathbb{V}_{ 1}\trans{\mathbb{V}}_{  1}+\mathbf{v}_{ \nt}\trans{\mathbf{v}}_{\nt}.
\end{equation}
It follows immediately that the $i$-th row of the matrix $\mathbf{v}_{ \nt}\trans{\mathbf{v}}_{\nt}$ is all-zero if $v_{\nt,i}= 0$. Due to $\mathbf{x}\in [0,1]^{\nt}$, we have $\kappa_{i}(\lambda) \in [0,1]^{\nt}$ in the case of $v_{\nt,i}= 0$ .

We only need to consider the indices $i$ such that  $v_{\nt,i}\neq 0$. Notice that $\kappa_{i}(\lambda)$ is an affine function, which {\color{black}is equal} to $0$ at
$
\lambda=\left(-\sum_{j=1}^{\nt-1} v_{j,i}s_j/\sigma_j\right)/ v_{\nt,i}
$
and equal to $1$ at
$\lambda=\left(1-\sum_{j=1}^{\nt-1} v_{j,i}s_j/\sigma_j\right)/ v_{\nt,i}.
$ It is clear that the inequality $
	0\le \kappa_{i}(\lambda) \le 1$ holds for all $  \lambda\in \left[\frac{1-\textnormal{sign}(v_{\nt,i})}{2v_{\nt,i}} -\sum_{j=1}^{\nt-1} \frac{v_{j,i}s_j}{v_{\nt,i}\sigma_j},
	\frac{1+\textnormal{sign}(v_{\nt,i})}{2v_{\nt,i}} -\sum_{j=1}^{\nt-1} \frac{v_{j,i}s_j}{v_{\nt,i}\sigma_j}
	\right]$
in the case of $v_{\nt,i}\neq0$. Thus, we conclude that
\begin{flalign}
	f_{\textnormal{min}}(\mathbf{s})=\max_{ {\color{black}i:\,} v_{\nt,i}\neq 0}\left\{ \frac{1-\textnormal{sign}(v_{\nt,i})}{2v_{\nt,i}} -\sum_{j=1}^{\nt-1} \frac{v_{j,i}s_j}{v_{\nt,i}\sigma_j} \right\}
\end{flalign}
and
\begin{flalign}
	f_{\textnormal{max}}(\mathbf{s})=\min_{ {\color{black}i:\,} v_{\nt,i}\neq 0}\left\{ \frac{1+\textnormal{sign}(v_{\nt,i})}{2v_{\nt,i}} -\sum_{j=1}^{\nt-1} \frac{v_{j,i}s_j}{v_{\nt,i}\sigma_j} \right\}.
\end{flalign}
It is noteworthy that $ f_{\textnormal{min}}(\mathbf{s})$ is a convex function, while $ f_{\textnormal{max}}(\mathbf{s})$ is concave. Especially, by the central symmetry of a cube, we can prove the following equality 
\begin{equation}\label{eq:68}
	f_{\textnormal{max}}(\mathbf{s})+f_{\textnormal{min}}(\widetilde{\mathbb{H}}\bm{1}-\mathbf{s})=\trans{\bm{1}}\mathbf{v}_{\nt}.
\end{equation}
Recall that $
	\mathcal{X}_{\widetilde{\mathbb{H}}\bm{1}-\mathbf{s}}$ consists of all points $ \lambda \mathbf{v}_{\nt}+\mathbb{V}_{ 1}\mathbb{B}_{11}^{-1}\left(\widetilde{\mathbb{H}}\bm{1}-\mathbf{s}\right)$ over $
	\lambda\in \left[  f_{\textnormal{min}}(\widetilde{\mathbb{H}}\bm{1}-\mathbf{s}),
	f_{\textnormal{max}}(\widetilde{\mathbb{H}}\bm{1}-\mathbf{s})\right]
$
and notice that
\begin{flalign}
	&\lambda \mathbf{v}_{\nt}+\mathbb{V}_{ 1}\mathbb{B}_{11}^{-1}\left(\widetilde{\mathbb{H}}\bm{1}-\mathbf{s}\right) \nonumber\\
	=&\lambda \mathbf{v}_{\nt}+\mathbb{V}_{ 1}\trans{\mathbb{V}_{ 1}}\bm{1}-\mathbb{V}_{ 1}\mathbb{B}_{11}^{-1}\mathbf{s} \nonumber \\
	=&\lambda \mathbf{v}_{\nt}+\left( \bm{1}-\left(\trans{\bm{1}}\mathbf{v}_{\nt}\right)\mathbf{v}_{\nt}\right)-\mathbb{V}_{ 1}\mathbb{B}_{11}^{-1}\mathbf{s} \label{eq:70} \\
	=&\bm{1}-\left(  \left( \trans{\bm{1}}\mathbf{v}_{\nt}-\lambda \right) \mathbf{v}_{\nt}  +\mathbb{V}_{ 1}\mathbb{B}_{11}^{-1}\mathbf{s}  \right),
\end{flalign}
where~\eqref{eq:70} follows from~\eqref{eq:62}. It is clear that $\mathcal{X}_{\widetilde{\mathbb{H}}\bm{1}-\mathbf{s}}$ is the reflection of $\mathcal{X}_{ \mathbf{s}}$ across the central point $\frac{1}{2}\bm{1}$. Hence, we know that the set $
	\mathcal{X}_{ \mathbf{s}}$ consists of points $\left( \trans{\bm{1}}\mathbf{v}_{\nt}-\lambda \right) \mathbf{v}_{\nt}  +\mathbb{V}_{ 1}\mathbb{B}_{11}^{-1}\mathbf{s}$ for all
	$ \lambda \in \left[  f_{\textnormal{min}}(\widetilde{\mathbb{H}}\bm{1}-\mathbf{s}),
	f_{\textnormal{max}}(\widetilde{\mathbb{H}}\bm{1}-\mathbf{s}) \right]
$, which immediately leads to~\eqref{eq:68}.

Based on Lemma~\ref{lemma:input}, we can alternatively let the channel input be the image of a random vector $\mathbf{S}$ under a deterministic mapping $\bm{\phi}:\mathcal{R}\left({\widetilde{\mathbb{H}}}\right) \to [0,1]^{\nt}$. Let $\mathscr{P}$ be a given probability measure of $\mathbf{S}$ on $\mathcal{R}\left({\widetilde{\mathbb{H}}}\right)$, and assume $\bm{\phi}$ to be Lebesgue measurable and satisfy $\bm{\phi}(\mathbf{s})\in \mathcal{X}_{\mathbf{s}}$ for each $\mathbf{s}\in \mathcal{R}\left({\widetilde{\mathbb{H}}}\right)$. We can write $\bm{\phi}(\mathbf{s})=\lambda(\mathbf{s}) \mathbf{v}_{\nt}+\mathbb{V}_{ 1}\mathbb{B}_{11}^{-1}\mathbf{s}$ readily, where $\lambda(\mathbf{s})\in \left[ f_{\textnormal{min}}(\mathbf{s}),\trans{\bm{1}}\mathbf{v}_{\nt}-f_{\textnormal{min}}(\widetilde{\mathbb{H}}\bm{1}-\mathbf{s})\right]$. Under the probability measure $\mathscr{P}$, the vector-valued expectation of $\phi(\mathbf{S})$ is given by 
\begin{flalign}\label{eq:25}
	\bm{\zeta}_{\! \scriptscriptstyle \mathscr{P}}^{\bm{\phi}}&=\int  \bm{\phi}(\mathbf{s}) \dd \mathscr{P}\\
	&=\left(\int  \lambda(\mathbf{s}) \dd \mathscr{P} \right) \mathbf{v}_{\nt}  +\int   \mathbb{V}_{ 1}\mathbb{B}_{11}^{-1} \mathbf{s} \dd \mathscr{P}.
\end{flalign}
Let $\mathcal{E}_{\!\mathscr{P}}=\left\{ \bm{\zeta}_{\! \scriptscriptstyle \mathscr{P}}^{\bm{\phi}} \right\} \subseteq [0,1]^{\nt}$, which contains real vectors $\bm{\zeta}_{\! \scriptscriptstyle \mathscr{P}}^{\phi}$ over all admissible mappings $\phi$. The convexity of $\mathcal{E}_{\!\mathscr{P}}$ can be easily proved since $\bm{\zeta}_{\! \scriptscriptstyle \mathscr{P}}^{\phi}$ is linear with respect to the mapping $\phi$, and thus the set $\mathcal{E}_{\!\mathscr{P}}$ must be a line segment parallel to $\mathbf{v}_{\nt}$.  Furthermore, note that
\begin{equation*}
	\int   f_{\textnormal{min}}(\mathbf{s}) \dd \mathscr{P} \le  \int  \lambda(\mathbf{s}) \dd \mathscr{P} 
	\le \trans{\bm{1}}\mathbf{v}_{\nt}-\int  f_{\textnormal{min}}(\widetilde{\mathbb{H}}\bm{1}-\mathbf{s}) \dd \mathscr{P} ,
\end{equation*}
where the lower bound is reached by letting $\lambda(\mathbf{s})=f_{\textnormal{min}}(\mathbf{s})$, while the upper bound is reached by $\lambda(\mathbf{s})=\trans{\bm{1}}\mathbf{v}_{\nt}-f_{\textnormal{min}}(\widetilde{\mathbb{H}}\bm{1}-\mathbf{s})$.
Then we obtain
\begin{flalign}
		\mathcal{E}_{\!\mathscr{P}}=\Big\{&\beta  \mathbf{v}_{\nt}+\mathbb{V}_{ 1}\mathbb{B}_{11}^{-1}  \mathbb{E} \left[\mathbf{S}\right]: \nonumber\\
	&\beta \in \big[\mathbb{E} \left[ f_{\textnormal{min}}(\mathbf{S})\right],\trans{\bm{1}}\mathbf{v}_{\nt} -\mathbb{E} \left[ f_{\textnormal{min}}(\widetilde{\mathbb{H}}\bm{1}-\mathbf{S})\right]\big] \Big\}.
\end{flalign}
Note that the vector $\bm{\alpha}=\mathbb{V}\trans{\mathbb{V}}\bm{\alpha}$, i.e., a linear combination of the basis $\mathbf{v}_1,\cdots,\mathbf{v}_{\nt}$
with the coordinate representation $\trans{(\trans{\mathbf{v}}_1\bm{\alpha},\cdots,\trans{\mathbf{v}}_{\nt}\bm{\alpha})}$. It is straightforward that a random vector $\mathbf{S}$ with the probability law $\mathscr{P}$ and $\supp\,  \mathbf{S} \subseteq \mathcal{R}\left({\widetilde{\mathbb{H}}}\right)$ is a feasible equivalent input if and only if $
\bm{\alpha}\in \mathcal{E}_{\!\mathscr{P}}$. Then, by comparing $\bm{\alpha}$ and $\mathcal{E}_{\!\mathscr{P}}$, we conclude that the probability law $\mathscr{P}$ of the equivalent input $\mathbf{S}$ is feasible if and only if \eqref{eq:equiv_ecmc} holds.
This is complete proof of Theorem~\ref{thm:nt-1}. \hfill$\blacksquare$

\vspace{-1.5em}
\section{Proof of Proposition~\ref{prop:maxvar}}\label{app:maxvar}
A case-by-case proof is given as what follows.
\begin{enumerate}
	\item $\alpha_{\nt}\ge \frac{1}{2}$. We conclude $\mathbf{x}^{\star}=\frac{1}{2}\bm{1}$ by showing the bivariate function $\psi(x,y)=\min \left\{x,y\right\}-xy$ is maximized when $x=y=1/2$.
	

	\item $\alpha_{\nt}< \frac{1}{2}$. For any feasible point $\mathbf{x}$, there exists a permutation matrix $\mathbb{P}_{\pi}$ such that elements of $\mathbf{z}=\mathbb{P}_{\pi} \cdot \mathbf{x}$ are in descending order. Note that the objective function $
		f_0(\mathbf{x})=\tilde{f}_0(\mathbf{z})=\trans{\mathbf{z}}\mathbb{G}_{\pi}\mathbf{z}-\trans{\mathbf{d}_{\pi}}\mathbf{z}$, where  $\mathbb{G}_{\pi}={\mathbb{P}_{\pi}}\trans{\mathbb{H}}\mathbb{H}\trans{\mathbb{P}_{\pi}}$, $
	\mathbf{d}_{\pi}\triangleq \trans{\left( \omega_1,\omega_2-\omega_1,\cdots,\omega_{\nt}-\omega_{\nt-1} \right)}$, and $
	\omega_i\triangleq{\left(\sum_{j=1}^i \trans{\mathbf{e}_j}\right)}\mathbb{G}_{\pi}\left(\sum_{j=1}^i \mathbf{e}_j\right)
	$. 
	Let $\left(\pi(1),\pi(2),\cdots,\pi(\nt)\right)$ be the permutation of $(1,2,\cdots,\nt)$ associated by $\mathbb{P}_{\pi}$. Consider the directional derivative that decreases the maximum of $\mathbf{x}$
	\begin{flalign}
		-\trans{\mathbf{e}_1}\nabla \tilde{f}_0(\mathbf{z})
		&=-\trans{\mathbf{e}_1} 
		\left(2\mathbb{G}_{\pi}{\mathbf{z}}+\mathbf{d}_{\pi}\right) \nonumber \\
		&=2\sum_{i=1}^{\nt}\trans{\mathbf{h}_{\pi(1)}}{\mathbf{h}_{\pi(i)}}z_i-\trans{\mathbf{h}_{\pi(1)}}{\mathbf{h}_{\pi(1)}} \nonumber \\
		&\le \trans{\mathbf{h}_{\pi(1)}}{\mathbf{h}_{\pi(1)}}\left(1-2x_{\pi(1)}\right),
	\end{flalign}
	which is nonpositive when the maximum coordinate $x_{\pi(1)}\ge \frac{1}{2}$.
	 Thus, we immediately know $\mathbf{x}^{\star} \preccurlyeq \min\left\{\alpha_{1},\frac{1}{2}\right\}\bm{1}_{\nt}$.
	
	Next, we consider another directional derivative that increases the minimum of $\mathbf{x}$ as follows
	\begin{flalign}
		\trans{\mathbf{e}_{\nt}}\nabla \tilde{f}_0(\mathbf{z})
		=\, &\trans{\mathbf{e}_{\nt}} 
		\left(2\mathbb{G}_{\pi}{\mathbf{z}}-\mathbf{d}_{\pi}\right) \nonumber \\
		\le \, & \sum_{i=1}^{\nt}\trans{\mathbf{h}_{\pi(\nt)}}{\mathbf{h}_{\pi(i)}}\left( 2z_i-1 \right) \nonumber \\
		\le \, & 0,
	\end{flalign}
	which implies that, in order to minimize the objective function $f_0$, the minimum coordinate should be maximized until failure to satisfy the constraints. Hence, we also conclude that $\bm{x}^{\star} \succcurlyeq \alpha_{\nt}\bm{1}$. \hfill$\blacksquare$ 
\end{enumerate}


\bibliographystyle{ieeetr}
\bibliography{./defshort1,./biblio1}

\addtolength{\textheight}{-68mm}

\end{document}